\documentstyle{mn}
\title[Angular Correlation Function and Hierarchical Moments of Faint Galaxies]{The Angular Correlation Function and Hierarchical Moments
of $\sim 70000$ Faint Galaxies to R=23.5} 
\author[N. Roche and S. A. Eales]{Nathan 
Roche$^{1,2}$ and Stephen A. Eales$^{1,3}$\\ 
$^1$Department of Physics and Astronomy,
      University of Wales Cardiff,
      P.O. Box 913,
      Cardiff CF2 3YB\\
{$^2$ \verb"ndr@astro.cf.ac.uk"}\hspace{8mm}   {$^3$ 
\verb"S.Eales@astro.cf.ac.uk"}\hspace{8mm}
}

\input{psfig.sty}

\begin{document}

\maketitle

\begin{abstract}

 We investigate the angular correlation function, $\omega(\theta)$, and
third and fourth-order hierarchical moments of a large sample of $\sim 70000$
galaxies of apparent magnitude  $18.5\leq R\leq 23.5$. The data consists of 47 red-band INT Wide Field Camera CCD images, forming two widely
separated fields of total areas 1.01 $\rm deg^2$ and 0.74 $\rm deg^2$.

 Galaxy clustering is detected with a high significance of $\sim 10\sigma$.
Over the  $R=21$ to $R=23.5$ range of magnitude limits, the angular correlation functions approximately follow $\theta^{-0.8}$ power-laws at $\theta>5$ arcsec, 
with amplitudes consistent with previous surveys and best-fitted by a luminosity evolution model in which galaxy clustering is approximately stable in proper co-ordinates ($\epsilon=0$). Assuming the redshift distribution from our pure
luminosity evolution model and the present-day galaxy correlation radii from
Loveday et al. (1995),  
we estimate the clustering evolution as $\epsilon=0.02_{-0.31}^{+0.48}$ for 
bth fields combined or $\epsilon=-0.29_{-0.27}^{+0.31}$ for the larger of the two fields which is thought to be the better in data quality.

 On the larger of our fields, $\omega(\theta)$ at $2\leq \theta \leq 5$
significantly exceeds a $\omega(\theta)\propto \theta^{-0.8}$ power-law of the amplitude measured at $\theta>5$ arcsec. If this excess of close pairs is due to interacting and merging galaxies, we estimate that it is consistent with the local fraction of galaxies in close ($<19 h^{-1}$ kpc) pairs combined with a merger/interaction rate
evolving as $(1+z)^m$ with $m=2.01_{+0.52}^{-0.69}$

 We derive the hierarchical moments $s_3(\theta)$ and
$s_4(\theta)$ from the counts-in-cells of the 
$18.5\leq R\leq 23.5$ galaxies. We find relatively steep slopes of approximately
$\theta^{-0.4}$ for $s_3(\theta)$ and $\theta^{-0.6}$ for $s_4(\theta)$, similar to those of the
hierarchical moments  
 of $17\leq B\leq 20$ galaxies in the EDSGCS (Szapudi et al. 1996)
at comparable physical scales.  
The $s_3(\theta)$ and
$s_4(\theta)$ of the $18.5\leq R\leq 23.5$ galaxies
show only a small reduction in normalization relative to those of the less deep EDSGCS survey,
consistent with the expectations from $N$-body simulations of the evolving mass distribution. This indicates there is little change in the mean linear or non-linear biasing of galaxies 
from $z\sim 0$ to $z\sim 1$, and hence appears to favour
 luminosity evolution for galaxies over  `transient starburst dwarf' models,  and
supports the interpretation of the stable galaxy clustering
 as evidence for a low $\Omega$ Universe.

\end{abstract}


\section{Introduction}

The clustering of galaxies on the sky is most often described in terms of the 
angular correlation function, $\omega(\theta)$, a projection onto the
sky plane of the three-dimensional two-point
correlation function $\xi(r)$. For galaxies of separation $r$ in proper 
co-ordinates, at redshift $z$, the two-point correlation function is often parameterized as
\begin{equation}
\xi(r,z)=({r\over r_0})^{-\gamma}(1+z)^{-(3+\epsilon)}
\end{equation}
where $r_0$ is the correlation radius, $\gamma$ the slope, and $\epsilon$ 
parameterizes the evolution of clustering (clustering stable in proper co-ordinates being $\epsilon=0$). The angular correlation
function will be a power-law of slope $\theta^{-(\gamma-1)}$ and amplitude given by the Limber's formula (e.g. Phillipps et al. 1978) integration of $\xi(r,z)$ over the galaxy redshift distribution $N(z)$.

Photographic galaxy surveys indicated that $\omega(\theta)$ approximately follows a $\theta^{-0.8}$ power-law,
with a break at very large scales, and an amplitude which decreases with increasing survey depth due to the effects of projection 
(e.g. Groth and Peebles 1977; Maddox et al. 1990).
During the early 1990s, a number of CCD surveys indicated that the
 $\omega(\theta)$ amplitude of faint ($B>23.5$) galaxies  falls well
below the scaling expected for stable clustering and a galaxy
$N(z)$ given by a non-evolving model (e.g. Efstathiou et al. 1991; Roche et al. 1993).
This was interpreted (Roche et al. 1993, 1996) as evidence that $N(z)$ at these magnitudes is
more extended than the non-evolving prediction, implying an  
evolutionary brightening of $L\sim L^*$ galaxies at higher redshifts.  
Alternatively, it could mean that the clustering of galaxies evolves rapidly
(i.e. $\epsilon\geq 1$) and so was much weaker at
 even moderate redshifts. Another possibility was that 
 that the faint blue galaxies at $B>23$ 
were a new population of blue dwarf galaxies at moderate
redshifts, with much weaker clustering than $L\sim L^*$ spirals (Babul and Rees 1992; Efstathiou 1995).
  
It was then found that the redder galaxies at these magnitudes 
gave a much ($\sim 0.5$ dex) higher
$\omega(\theta)$ amplitude than the bluer galaxies (Neuschaefer et al. 1995;
Roche et al. 1996). This
would argue against a strong evolution of galaxy clustering as an explanation of the low $\omega(\theta)$ amplitudes, but
 would be consistent with $L^*$ evolution models, where the extended tail of $N(z)$ would consist of blue star-forming giant galaxies. However, it would also be consistent with intrinsically weakly clustered blue dwarf galaxies at lower redshift.

 Soon afterwards, the importance of $L^*$ evolution was established directly  when deeper redshift surveys with the Keck telescope (e.g. Cowie, Songaila and Hu 1996; Steidel et al. 1996) and Hubble Deep Field photometry (e.g. Madau et al. 1996; Bershady et al. 1997)  
identified many galaxies at $z\sim 1$--3 and confirmed there is a brightening of the galaxy $L^*$ at these redshifts.

As the luminosity evolution is becoming better determined by other data, the emphasis of studying faint galaxy clustering is changing from the use of $\omega(\theta)$ as a probe of $N(z)$ towards 
a more detailed investigation of the evolution of clustering and large-scale structure, 
the clustering properties of different types of galaxy, and the possible 
effects of interactions between galaxies. In particular, there is still much 
to learn about the rate of clustering evolution, $\epsilon$, which is sensitive to the cosmological parameters and also to any 
biasing of the luminosity distribution relative to the underlying mass distribution. Models 
(e.g. Col\'{i}n, Carlberg and Couchman 1997) and direct observations from redshift surveys 
(e.g. Le F\`{e}vre et al. 1996; Shepherd, Carlberg and Yee
 1997; Carlberg et al. 1997)
generally span a wide range of at least $0\leq \epsilon\leq 1.5$, although models in which the bias factor increases rapidly with redshift (e.g.  Matarrese
et al. 1997) may predict galaxy clustering evolution closer to a comoving
model with $\epsilon=\gamma-3\simeq -1.2$.

The first deep CCD surveys covered small ($<0.1$
 $\rm deg^2$) areas of sky,
on which the detection of galaxy clustering was only $\sim 3\sigma$,
but with large format CCDs, such as those in the Wide Field Camera installed on the Isaac Newton Telescope in 1997, it becomes possible to survey 
$>1$ $\rm deg^2$ areas and study faint galaxy clustering
in more detail. In this paper we present a study of galaxy clustering to
a limit $R=23.5$ on two
fields with  total areas 1.01 $\rm deg^2$ and 0.74 $\rm deg^2$.
The larger fields will greatly reduce the `integral constraint' correction (see Section 4), and the increase in sample size will provide better constraints on the clustering evolution.

 Previous, less deep, surveys  suggest
that $>1$ $\rm deg^2$ areas can provide more than just the $\omega(\theta)$ amplitude. For example, Infante and Pritchet (1995), using a $2 \rm deg^2$ 
mosaic of photographic plates, reported
some flattening of the slope of $\omega(\theta)$ between $R\simeq 21$ and their limit of $R\simeq 23$, while Infante, de Mello and Mentaneau (1996), using a similar area made up of short-exposure CCD images, found the  $R<21.5$ galaxy $\omega(\theta)$ to be significantly higher in amplitude at $2<\theta<6$ arcsec than at larger scales. With a large area,
it may also be possible to measure the higher-order correlations functions, which are related to the higher-order moments (skewness, kurtosis etc.) of the galaxy counts-in-cells, in the same way that $\omega(\theta)$ is related to the variance. These statistics provide additional information on the nature of galaxy clustering but until now have been studied only
at  much shallower limits (Gazta\~naga 1994; Szapudi, Meiksin and Nichol 1996). 

We assume $H_0=50$ km $\rm s^{-1} Mpc^{-1}$ and $q_0=0.05$ throughout, but
give some quantities in units of $h=H_0/100$ km $\rm s^{-1} Mpc^{-1}$. 
In Section 2 of this paper we describe the observational data and its reduction and calibration, and in Section 3 the number counts of the galaxies and stars. Section 4 describes the 
analysis of the galaxy $\omega(\theta)$, and Section 5 presents the results.
Section 6 describes models of
galaxy evolution and clustering and compares these with our observations and other data. In Section 7 we discuss $\omega(\theta)$
at small scales of a few arcsec and its use in investigating merger rate evolution. In Section 8
we investigate the higher-order moments of the sample and compare with
results from less deep surveys. Section 9 discusses in detail the implications of all these results.    

\section {Data}

\subsection {Observations}

Our data were acquired on the nights of 4th and 5th June 1997, on a service 
run of the recently installed Wide Field Camera (WFC), at the prime focus of the 2.5 metre Isaac Newton Telescope on La Palma. At that time the WFC was fitted with a
$2\times 2$ array of Loral CCDs, each with $2048\times 2048$ pixels.  In this
configuration the pixelsize corresponds to 0.37 arcsec. The WFC should then have
provided a total area of approximately 638 $\rm arcmin^2$ with each pointing,
but one of the four CCD chips was nonfunctional and another
suffered a severe loss of sensitivity towards the edges, reducing the usable
area to $\sim 440$ $\rm arcmin^2$. 

The WFC imaged 21 positions, 12 arranged in a $3\times 4$
grid referred to as field `e', and 9 in a $3\times3$ grid referred to 
field `f'. Field `e' is centred at R.A. 14:00 hours, declination zero, which is galactic longitude $337.01^{\circ}$ and latitude $+58.26^{\circ}$, and field 
`f' is centred at R.A. 21:00 hours, declination $-10{^\circ}$, which is
 galactic longitude $38.71^{\circ}$ and latitude $-33.06^{\circ}$. Each of the 21 pointings was exposed twice for 900 seconds, 
 with the WFC Harris Red filter. For photometric
calibration, 5 second exposures were taken of several fields containing standard stars from the catalog of Landolt (1992).

Some further sections of the data were rendered unusable by autoguider malfunctions and a period of poor seeing. We were able to use a total of 47 CCD frames (27 in field `e' and 20 in field `f') from 16 pointings of the WFC at co-ordinates given in Table 1. 
\begin{table}
\caption{Positions (equinox 2000.0) of the 16 pointings of the WFC 
and estimates of the resolution on the images}
\begin{tabular}{lccc}
\hline
Field name    & R.A. & Dec. & FWHM (arcsec) \\
\smallskip
e1  & 14:00:00.00 & +00:00:00.3 &  1.43 \\           
e2  & 13:59:59.98 & +00:25:39.9 &  1.41 \\
e4  & 14:00:00.00 & -00:51:19.7 &  1.51 \\
e7  & 14:01:38.05 & +00:00:00.3 &  1.70 \\
e8  & 14:01:37.94 & -00:24:29.7 &  1.73 \\
e9  & 14:01:37.84 & -00:49:01.3 &  1.62 \\
e10 & 13:58:22.10 & -00:00:00.1 &  1.51 \\
e11 & 13:58:22.07 & +00:24:29.3 &  1.63 \\
e12 & 14:01:38.16 & +00:24:29.3 &  1.88 \\
f1  & 21:00:00.09 & -10:00:00.2 &  1.92 \\
f2  & 20:59:59.97 & -10:24:29.9 &  1.89 \\         
f4  & 21:01:39.00 & -09:35:29.8 &  1.72 \\
f6  & 21:01:38.98 & -10:24:25.7 &  2.19 \\          
f7  & 20:58:21.07 & -09:35:29.2 &  1.68 \\
f8  & 20:58:21.01 & -09:59:58.5 &  1.55 \\
f9  & 20:58:21.03 & -10:24:28.7 &  1.78 \\
\hline
\end{tabular}
\end{table}
\subsection{Reduction and source detection}
The initial processing of the data was carried out using {\sevensize IRAF}. For each of the three functional CCD chips, a bias frame was produced and subtracted from all images. One of the three CCD chips showed a significant falloff in sensitivity towards the edges. We trimmed these images approximately where the sensitivity fell to 0.64 of its central value, reducing the area by $\sim 25\%$.
For each chip, median flat fields were generated from the data
using {\sevensize IRAF} `ccdcombine', by taking the median of all 32 exposures on the positions in Table 1, rejecting those pixels containing bright objects.
All images were then divided by the respective normalized flat fields.

The two exposures at each position were added using {\sevensize IRAF} `ccdcombine',
using the `crrej' option (rejecting the higher of the two values if discrepant by $>3\sigma$) to remove the majority of cosmic rays.  Sky subtraction was performed by  fitting high-order spline surfaces to the sky background on each image, using {\sevensize KAPPA} `surfit', and subtracting these from the data.  The mean sky background intensity (using the calibration in Section 2.3) was
$20.77\pm 0.02$ $R$ mag $\rm arcsec^{-2}$ on field `e' and $20.70\pm 0.02$ $R$ mag $\rm arcsec^{-2}$ on field `f'.

   The Starlink {\sevensize PISA} (Position, Intensity and Shape Analysis) package, developed
 by M. Irwin, P. Draper and N. Eaton, was used to detect and catalog the
objects, with the  chosen 
detection criterion that a source must exceed an intensity threshold of
$1.5\sigma$ above the background noise ($\sigma$ being separately determined by
{\sevensize PISA} for each image) in at least 6 connected
pixels. The mean detection threshold was $25.72\pm 0.01$ $R$ mag $\rm arcsec^{-2}$ on field `e' and $25.63\pm 0.02$ $R$ mag $\rm arcsec^{-2}$ on field `f'. {\sevensize PISA} was run with deblend (objects connected at the detection threshold may be split if they
are separate at a higher threshold) and the `total magnitudes' option, which estimates total intensity above the background by fitting an elliptical aperture to each individual detection and performing a `curve-of-growth' analysis.

We excluded data from circular `holes' (radius 15 to 100 arcsec) around many of the bright, saturated stellar images  in order to remove spurious noise detections. Radial profiles were fitted 
to several non-saturated stars on each image using `pisafit', giving the FWHM estimates in Table 1. Evidently, the seeing was less than
ideal during the period of observation,  resulting in a FWHM averaging $1.60\pm 0.04$ arcsec on field `e' and $1.82\pm 0.07$ arcsec on field `f', with much variation between the individual images.

Star-galaxy separation was performed using plots of central against total intensity, normalized to the ratio from the fitted stellar profile.
Figures 1 and 2 show the star-galaxy separation plots for one CCD frame from each of the two fields.
The stellar locus remains separate from the galaxies only to $R\sim 21$--21.5;
so detections fainter that this are all classified as galaxies. 
Even at $20.5<R<21.5$ the reliability of star-galaxy separation appears to vary between images. Hence we only assume the separation is reliable to $R=20.5$, and later (Sections 3 and 5) apply corrections to our results for the effects of star contamination at $R>20.5$. 
\begin{figure}
\psfig{file=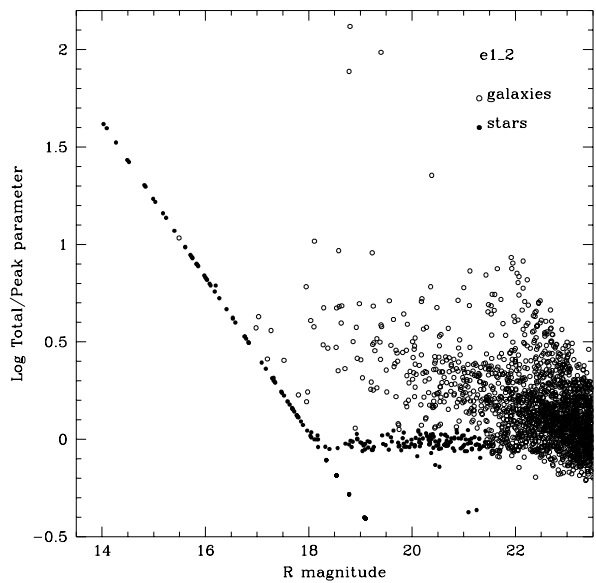,width=85mm}
\caption{The total/peak intensity parameter used for star-galaxy separation, plotted against $R$ magnitude, for detections on one CCD frame within field `e', showing their classification as stars or galaxies.
The rise in the stellar locus at $R>18$ is the result of saturation of the 
CCD chips imposing a upper limit on the measured peak intensity.}
\end{figure}
\begin{figure}
\psfig{file=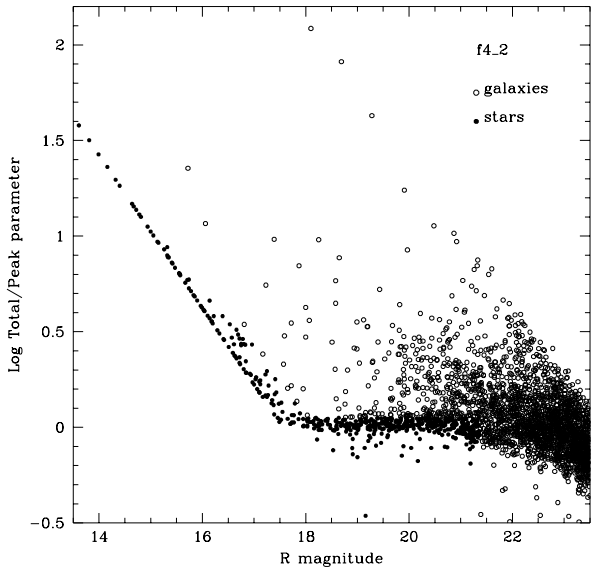,width=85mm}
\caption{As Figure 1 for one image within field `f'.}
\end{figure}
\subsection {Calibration}

For photometric calibration, the 900s exposures were interspersed with short exposures of three fields
containing standard stars (about 5 per CCD chip per field) from the catalog of Landolt (1992). After flat-fielding the standard star images, the intensities of the standard stars were measured using {\sevensize IRAF} aperture photometry.

For each observation of a standard
star, the photon count rate $C$ will
relate to the catalogued red magnitude $R$ and colour $V-R$ as 
\begin{equation}
-2.5 \log (C)=R + r_1 +r_2X +r_3(V-R) 
\end{equation}
where $r_1$ is the zero-point outside the atmosphere, $r_2$
the atmospheric extinction at the zenith, $X$ the airmass
of each observation is and $r_3$ a 
colour-correction term expressing the difference between our passband and the Cousins $R$ band of
Landolt (1992). As the standard star observations were too few in number and covered too small a range of airmass to accurately measure $r_2$, we estimated $r_2=0.062$ by integrating a 
tabulated measurement of the extinction curve at La Palma over the $R$ passband.

 The three CCD chips were calibrated separately. The
colour term $r_3$ was consistent with zero
within the statistical errors, so  the standard star observations
were fitted with
\begin{equation}
-2.5 \log (C)=R + r_1 + 0.062X 
\end{equation}
giving
$r_1=-25.185\pm0.012$, $-25.135\pm 0.005$ and $-25.369\pm 0.004$ for the
three CCD chips.    
The photon counts for each detected source could then 
be converted to $R$ magnitudes at the zenith with
\begin{equation}
R=-2.5 \log (C)-(r_1+0.062X)
\end{equation}
with $r_1$ and $X$ appropriate to the chip and airmass of observation. 

A further correction is applied for Galactic dust extinction. The Caltech NED database gives $B$-band extinctions of 0.10 and 0.19 magnitudes for objects near the centres of fields `e' and `f' respectively, approximately in proportion to the Galactic HI column densities. Using the extinction curve of Mathis (1990), we estimate $A_R=0.65 A_B$, and apply corrections of
 $\Delta(R)=-0.065$ and $\Delta(R)=-0.124$ to magnitudes from fields `e' and `f' respectively.

\subsection {Astrometry}

Astrometric transforms were derived using of the APM database
at the Institute of Astronomy in Cambridge. Field `e' lay within the measured UKST plate
F865 and field `f' within the measured UKST plate F742. We identified 12 catalogued stars on each of the 47 CCD images, and using their R.A. and Dec. fitted
astrometric transforms (with residuals typically only $\sim 0.5$ arcsec) using the {\sevensize IRAF} `pltsol' routine. The pixel co-ordinates of all detections were then converted to R.A. and Dec. The catalogs of detections on the  47 individual images could then be combined 
into two composite catalogs for the `e' and `f' fields.

 Most of the 16 exposures overlap with adjacent pointings by
a few arcminutes.
 For the areas of overlap, we adopted the positions and magnitudes derived from the image with the better seeing (Table 1) when combining the detection catalogs.  
Figures 3 and 4 show the R.A and Dec positions of the brighter galaxies in the
composite catalogs for fields `e' and `f'. In calculating angular 
separations between galaxies we assume 
$\Delta(\theta_X)/\Delta({\rm R.A})$ to be unity on field `e' and $\cos(10^{\circ})$ on field `f', neglecting the very small ($<0.02\%$) effect of sky curvature within each field.
After taking into account the overlap regions and holes, the areas 
covered by usable data amount to  1.012 $\rm deg^2$ for field `e' and 0.740 $\rm deg^2$ for field `f'. 
\begin{figure}
\psfig{file=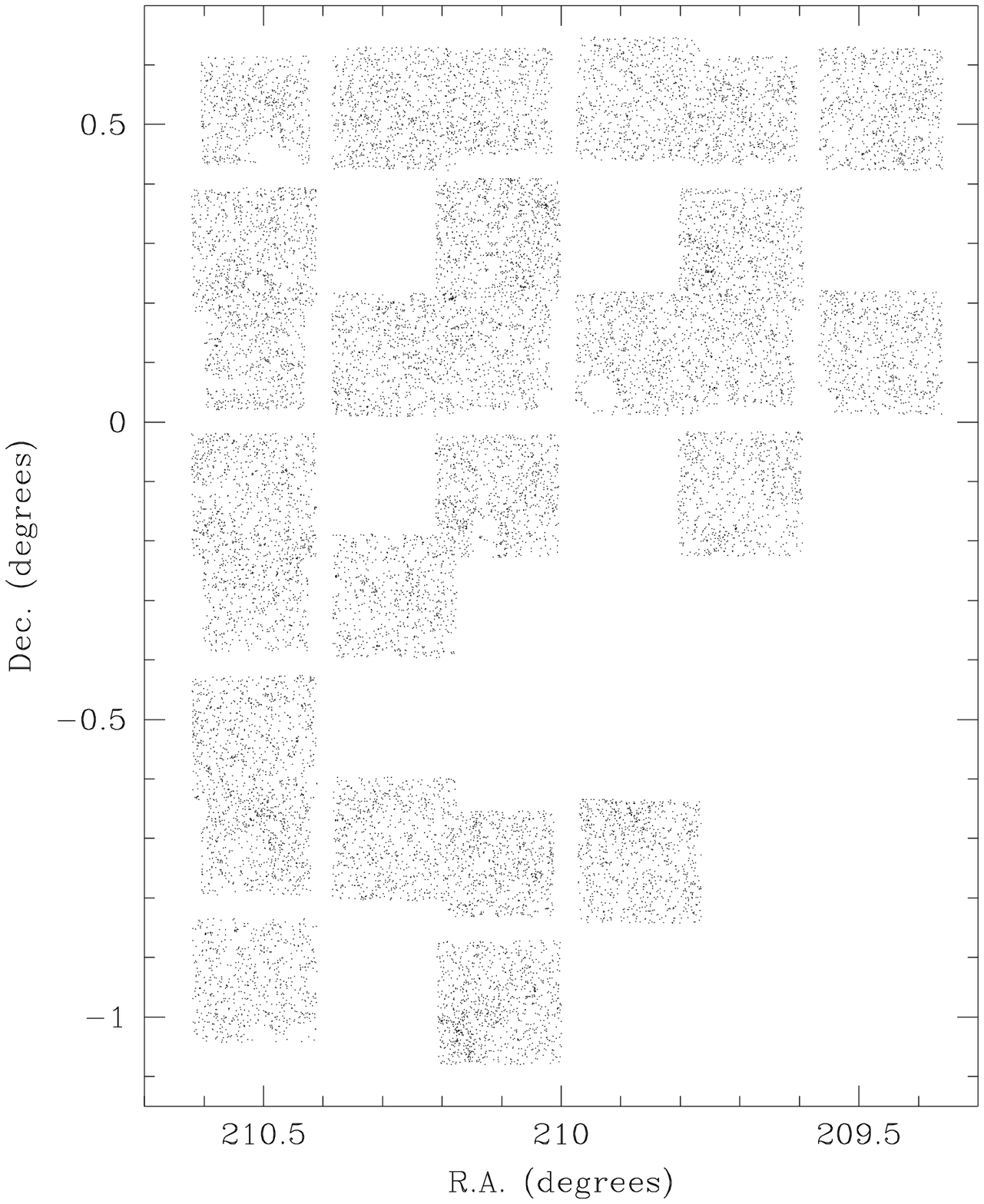,width=105mm}
\caption{The positions of detections classed as $18.5<R<22.5$ galaxies on the 27
CCD frames comprising field `e'} 
\psfig{file=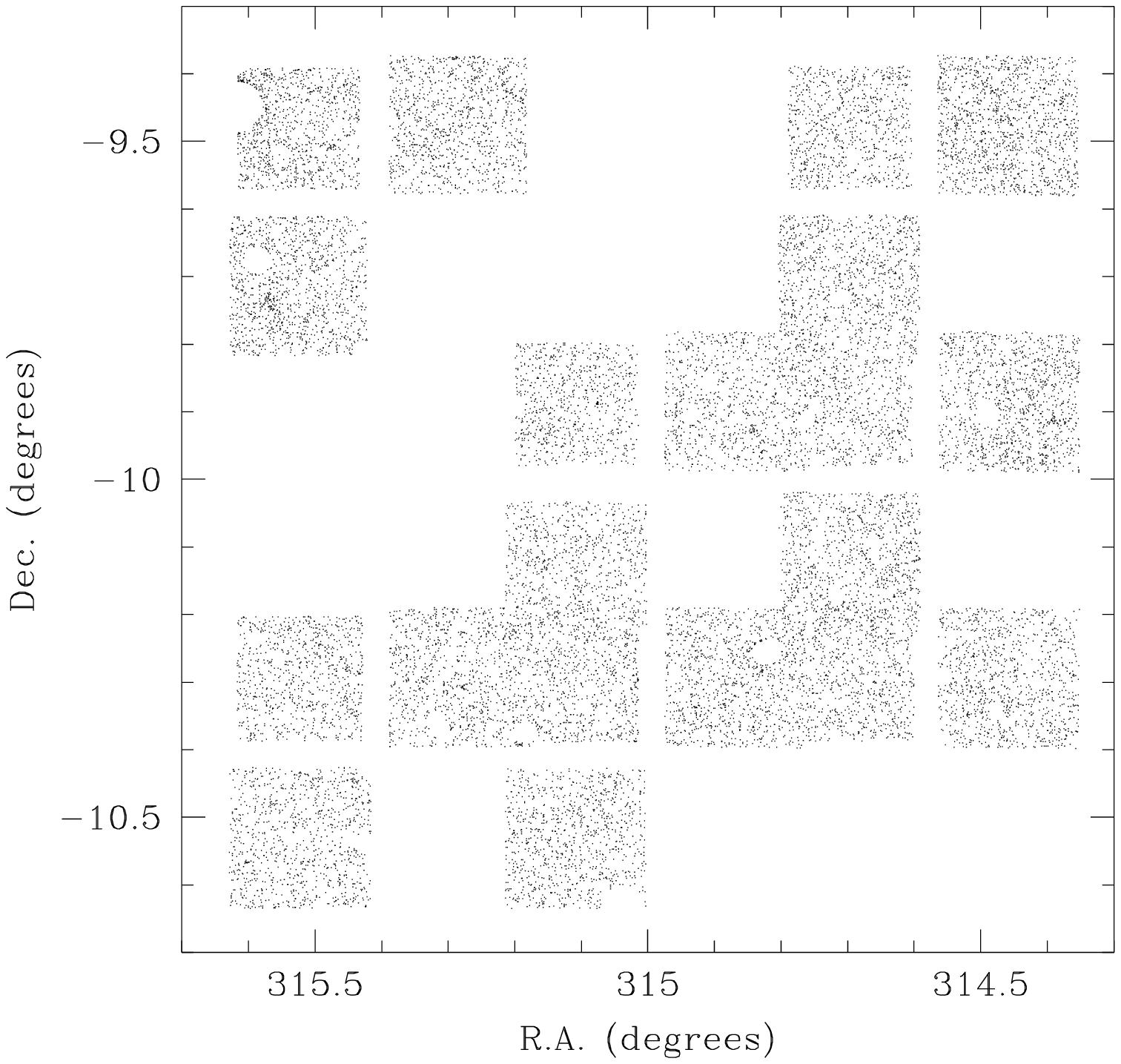,width=105mm}
\caption{The positions of detections classed as $18.5<R<22.5$ galaxies on the 20 CCD frames comprising field `f'} 
\end{figure}
\subsection {Photometric Matching in Overlap Areas}
The CCD frame overlap areas were made use of in two ways -- firstly, to estimate the photometric errors for individual faint galaxies by matching the two
 detections of the same objects and comparing the magnitudes. The mean scatter between the two detections gave errors as
$\sigma(R)=0.14$ mag for $R<22$ detections, 0.22 mag at $22<R<23$ and 0.24 mag at $23<R<24$. 

Secondly, by calculating a mean offset between the magnitudes of the same objects in each overlap, we can check for photometric shifts between the exposured, typically to $\sim 0.05$ mag accuracy. Significant offsets were seen for  
three exposures only -- e7, e8 and f6 were estimated to have suffered losses of   $0.18\pm 0.06$ mag, 
$0.41\pm 0.09$ mag and $0.09\pm 0.04$ mag respectively, in comparison to the rest of the data, presumably as a result of poor atmospheric
conditions. Appropriate corrections ($-0.18$, $-0.41$ and $-0.09$ mag) were therefore applied to all magnitudes from these fields.
The overlap areas will reveal only the worst of the photometric offsets, and, inevitably, there will be some remaining 
 scatter in the exposure zero-points -- this is investigated further in Section 5.3.

\section {Galaxy and Star Counts}
Figure 5 shows the differential number counts of objects classed as galaxies on
the `e' and `f' fields (error bars are derived from the scatter in sufface density between the individual CCD frames making up each field), together with galaxy counts from some previous CCD surveys in similar passbands, and the PLE and non-evolving models described in Section 6. Comparison with deeper galaxy counts suggests our detection of galaxies is virtually complete to $R=23.5$, but becomes significantly ($\sim30\%$ on field `e', $\sim 45\%$ on field `f') incomplete at $23.5<R<24$, before our counts turn over at $R>24$. As incompleteness can unpredictably affect measures of clustering,
our analysis is limited to to the $R\leq 23.5$ galaxies.

\begin{figure}
\psfig{file=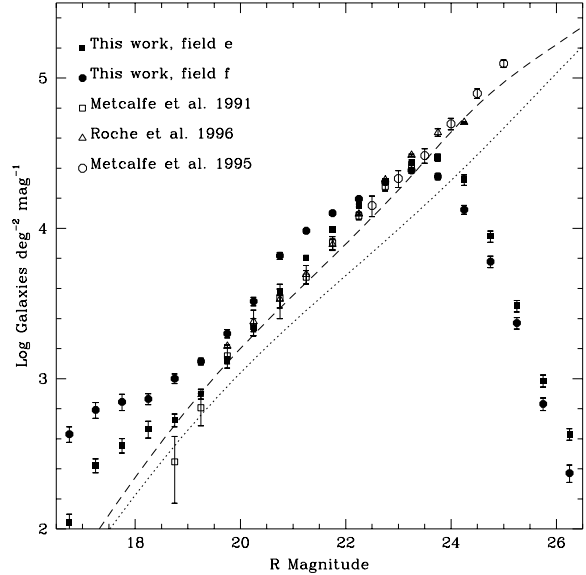,width=85mm}
\caption{Differential galaxy number counts, in 0.5 mag intervals of $R$ magnitude, for our two survey fields compared to previous CCD surveys in very similar passbands
and the predictions of our PLE (dashed) and non-evolving (dotted) models.} 
\end{figure}

\begin{figure}
\psfig{file=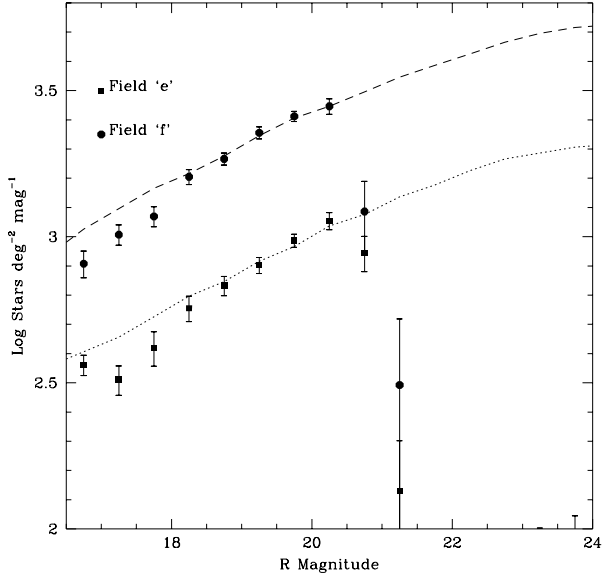,width=85mm}
\caption{ Differential star number counts in 0.5 mag intervals of $R$ magnitude, with star count models (dotted line for field `e', dashed line for field `f') from Reid et al. (1996), 
normalized to the observed count at $18<R<20.5$.} 
\end{figure}

At $18.5\leq R\leq 23.5$, the field `e' galaxy counts are close to both previous observations and the PLE model, whereas the field `f' counts are significantly 
higher at $20.5<R<22.5$. The galaxy counts also  show an excess above the models at bright magnitudes of $R>18.5$, on both fields but much more so on field `f'. These discrepancies are the result of star-contamination. 

We estimate the contamination by stars faintward of the limits of
star-galaxy separation ($R\simeq 20.5$--21) using the Galactic star count models from Reid et al. (1996). The $R$-band models for Galactic latitudes $b=60\degr$ and $b=45\degr$ are adopted for fields `e' and `f' respectively and normalized to fit our
observed star counts (figure 6) at $18.0<R<20.5$. The fitted normalizations corresponded 
to $1087\pm 76$ stars $\rm mag^{-1}deg^{-2}$ on field `e' and 
$2791\pm 136$ stars $\rm mag^{-1}deg^{-2}$ on the lower Galactic latitude field `f', with errors from the scatter in star counts
between the CCD frames within each field.

 We then  (i) assume stars are classified correctly to
$R=20.5$, (ii) estimate the number of $R>20.5$ stars by summing the
 normalized model star-counts from $R=20.5$ to the magnitude limit being considered, and (iii) subtract the number of $R>20.5$ detections already classified as stars. To $R=23.5$, this gives the total number of contaminating stars as $4306\pm 301$ on field `e' and $8433\pm 411$ on field `f'.

Brightward of $R\sim 18$--18.5 the star counts from both fields fall below the plotted models, approximately mirroring the excess seen in the galaxy
counts. Saturation of the brighter star images in our data evidently results in significant numbers becoming misclassified as galaxies. As we are interested in the clustering of much fainter galaxies, we remove this problem simply by excluding all $R>18.5$ detections from our analysis. This leaves a total of 80293 objects in our dataset classed as $18.5\leq R\leq 23.5$
galaxies, of which we estimate  $12739\pm 509$ are stars, leaving
$67554\pm 509$ genuine faint galaxies.
\section {Calculating $\omega(\theta)$}
The angular correlation function $\omega(\theta)$ was calculated separately for fields `e' and `f'. For field `e' we used all galaxies brighter than a series of limits from $R=21.0$ to $R=23.5$, and on field `f' used the same faint limits but also excluded $R<19.0$ objects.
To calculate $\omega(\theta)$ for a field containing $N_{gal}$ galaxies in the chosen magnitude range, the separations
of all ${1\over2}N_g(N_g-1)$ possible galaxy-galaxy pairs were counted
in bins of width $\Delta ({\rm log}~\theta)=0.2$, giving the function $N_{gg}(\theta_i)$

For each field $N_r=100000$ points were placed at random over the area
covered by real data. i.e. avoiding any gaps or holes.
 The separations of the $N_r N_g$ galaxy-random pairs, taking the real galaxies as the centres, were similarly binned, giving
$N_{rg}(\theta_i)$
and likewise the separations of the ${1\over2}N_r(N_r-1)$ random-random
pairs were counted to give $N_{rr}(\theta_i)$.
 
Defining $DD=N_{gg}(\theta_i)$, and $DR$ and $RR$ as the galaxy-random
and random-random counts normalized to have the same summation over all 
$\theta$ as $DR$, i.e.
$DR={(N_g-1)\over 2N_r} N_{gr}(\theta_i)$ and
$RR={N_g(N_g-1)\over N_R(N_r-1} N_{rr}(\theta_i)$,
 we might then estimate $\omega(\theta)$ for each bin simply as
(Roche et al. 1993, 1996)
\begin{equation}
\omega(\theta_i)={DD\over DR} -1
\end{equation}
However, Landy and Szalay (1993) presented a new estimator of $\omega(\theta)$, said to give the smallest possible statistical errors, and to optimally remove edge effects (Bernstein 1994), of the form
\begin{equation}
\omega(\theta_i)={DD-2DR+RR\over RR}
\end{equation}
 Ratcliffe et al. (1998) compared these clustering estimators using simulated galaxy catalogs, and confirmed that the Landy and Szalay (1993) estimator was significantly more accurate than that in equation (5) for the correlation functions at larger separations, so it will be used for the analysis of this paper.

Figures 7 and 8 show $\omega(\theta)$ calculated with equation (6) for fields
`e' and `f' respectively. The error bars in $\omega(\theta)$ estimates can be much larger than the Poisson error ($DD^{-0.5}$) and depend in a complex fashion on the field 
geometry, galaxy surface density and higher-order correlation functions. We estimate the error bars using a `jackknife' method, in which $\omega(\theta)$
is recalculated 10 times for each field, each time excluding just one of the
CCD frames (randomly chosen) from the analysis. As these subsamples will be smaller than the full fields by only one part in 27 (field `e') or 20
(field `f'), the effects of altering the field geometry will be minimized. For field `e', the error on $\omega(\theta)$ for the full field of 27 images was estimated by multiplying the
scatter in $\omega(\theta)$ between these 10 subsamples by $\sqrt 26$ (as only one part in 26 of the data in changed from one subsample to the next) $\times \sqrt {26\over 27}$ (as the full field contains 27 rather than 26 images)
$=5.00$. For field `f', the error for the full field of 20 images  
 was similarly estimated by multiplying the scatter between the subsample $\omega(\theta)$ images by $\sqrt 19 \times \sqrt {19\over 20}=4.25$.

If the real galaxy $\omega(\theta)$ is of the form $A\theta^{-0.8}$, the observed $\omega(\theta)$ will follow the form $\omega(\theta)=
A(\theta^{-0.8} - C)$, with amplitude $A$ (defined here at a one-degree
separation), and a negative offset $AC$ known as the integral
constraint. This offset results from the restricted area of the
observation, and can be estimated by doubly
integrating an assumed true $\omega(\theta)$ over the field area $\Omega$, i.e. 
\begin{equation}
AC={1\over \Omega^2} \int\int \omega(\theta) d\Omega_1 d\Omega_2
\end{equation}
This calculation can be done numerically using the random-random
correlation. Assuming $\omega(\theta)=A\theta^{-0.8}$,
\begin{equation}
C={\sum N_{rr}(\theta) \theta^{-0.8}\over \Sigma N_{rr}(\theta)}
\end{equation}

We then fit $\omega(\theta)$ with a function  
$\omega(\theta)= A(\theta^{-0.8} - C)$, but find that, at least on field `e', the two $\omega(\theta)$ points  at $2<\theta<5$ arcsec lie significantly above  a $\theta^{-0.8}$ power-law fitted at larger scales. 
Infante et al. (1996) had previously reported that the $\omega(\theta)$ of a large sample of $R\leq 21.5$ galaxies appeared to retain a  $\theta^{-0.8}$ slope at $2< \theta <6$ arcsec but was increased in amplitude by a factor 1.8
relative to the $\omega(\theta)$ at $\theta>6$ arcsec.
This result, indicating an excess of close pairs, was interpreted as the result of galaxy interactions and mergers.

In order to quantify the effects of interactions in our deeper survey, and at the same time minimize their influence on the measurement of larger scale galaxy clustering, we 
fit our $\omega(\theta)$ with two $\theta^{-0.8}$ power-laws, one at  $\theta>5$ arcsec to derive an amplitude $A$, another at  $2<\theta<5 $ arcsec to derive a `small scale amplitude' $A_{s}$ (at $\theta<2$ arcsec many pairs will be unresolved so the $\omega(\theta)$ cannot be used). 

The integral constraint then becomes
$AC_1 +A_{s}C_2$ where $C_1$ and $C_2$ are, respectively, the summations of equation (9) at all
angles $\theta>5$ arcsec and at
$2<\theta<5$ arcsec only. For field `e', $C_1=1.799$ and $C_2=0.00441$ and
for field `f', $C_1=2.024$ and $C_2=0.00517$. As the integral constraint
is greatly dominated by the large-scale $\omega(\theta)$, so we can neglect the
contribution from any small-scale excess and  assume the integral constraint to be $A(C_1+C_2)$. We estimate $A$ by least-squares fitting 
$A(\theta^{-0.8}-(C_1+C_2))$ to the observed $\omega(\theta)$ from 5 arcsec to  7962 arcsec on field `e' and 5024 arcsec
on field `f'. Once $A$ has been determined, giving the integral constraint,
$A_s$ is determined by fitting $A_s\theta^{-0.8}-A(C_1+C_2)$, to the two bins
at $2\leq \theta<5$ arcsec.

 The fitted functions, plotted on Figures 7 and 8, fit the field `e'
$\omega(\theta)$ at all separations, but the field `f' $\omega(\theta)$ 
shows some evidence of an excess above the power-law at ${\rm log}~\theta\geq -1.2$  -- we discuss a possible explanation in Section 5.3.

For comparison, we also calculate `individual frame' $\omega(\theta)$, in which each CCD frame is treated as an independent field for
which $\omega(\theta)$ calculated using equation n above with $N_r=20000$, and the results averaged for the 27 frames of field `e' and the 20 frames of field `f'. The integral constraints are larger for the small fields of invididual
frames, averaging $C=8.28$. In Section 5.3 these estimates are compared with the full-field $\omega(\theta)$.

\section {$\omega(\theta)$ Results} 
\subsection{Full-field estimates}
Table 2 gives the fitted $A$ and $A_s$ amplitudes for the two mosaiced
fields, for galaxies from $R=18.5$ to a series of faint limits from $R=21.0$ to $R=23.5$.
The errors were estimated by fitting amplitudes to our
measurements of $\omega(\theta)$ from datasets with one image excluded, and
 multiplying the scatter between these amplitudes by 5.00 for field `e' and by 4.25 for field `f'. 
\begin{table}
\caption{Observed $\omega(\theta)$ amplitudes at $\theta>5$ arcsec ($A$)
and at $2<\theta<5$ arcsec ($A_s$), in units of $10^{-4}$ at $1^{\circ}$, for the galaxies, numbering $N_{gal}$, with $R$ magnitudes from $R=18.5$ to a series of faint limits.}
\begin{tabular}{lccc}
\hline
$R$ limit & $A$ & $A_s$ & $N_{gal}$ \\
    & \multispan{3} \hfil Field `e'  \hfil \\
21.0 & $38.93\pm 4.97$ & $55.44\pm 5.57$ &  4395 \\
21.5 & $27.00\pm 2.00$ & $40.15\pm 5.58$ & 7643\\
22.0 & $20.20\pm 1.32$ & $38.61\pm 3.31$ & 12665  \\
22.5 & $17.66\pm 1.54$ & $33.31\pm 2.94$ & 19890 \\
23.0 & $15.73\pm 1.85$ & $27.28\pm 2.49$ & 30284 \\
23.5 & $16.07\pm 2.08$ & $23.74\pm 1.89$ & 44307  \\
\smallskip
    & \multispan{3} \hfil Field `f'  \hfil \\
21.0 & $26.52\pm 2.77$ & $27.50\pm 6.71$ & 5274 \\
21.5 & $15.29\pm 2.59$ & $12.97\pm 2.77$ & 8877 \\
22.0 & $13.06\pm 1.81$ & $11.97\pm 2.36$ & 13567 \\
22.5 & $12.00\pm 2.11$ & $13.70\pm 1.87$ & 19415 \\
23.0 & $12.59\pm 2.18$ & $14.86\pm 2.41$ & 26958 \\
23.5 & $13.83\pm 2.54$ & $16.60\pm 2.50$ & 35986 \\

\hline
\end{tabular}
\end{table}

However, the fitted amplitudes are likely to be underestimates due to 
the effects of star contamination. A fraction of (randomly distributed) stars $f_s$ contaminating the galaxy sample will dilute the observed $\omega(\theta)$ at all angles by a factor of
$(1-f_s)^{-2}$. Table 3 gives estimates of $f_s$ in our data from  the fitted star count models of Section 3.4, and $\omega(\theta)$  amplitudes
$A$ and $A_s$ corrected for star contamination by multiplying by
$(1-f_s)^{-2}$, with errors which include the (small) contribution from the uncertainty in $f_s$, added in quadrature to the $\omega(\theta)$ errors.

These results show that, firstly, our detection of clustering 
is of high significance, e.g. at $R\leq 23.0$ an estimated $8.5\sigma$ on field
`e' and $5.8\sigma$ on field `f', or $10.3\sigma$ overall.
Secondly, the  two  fields give reasonably consistent $\omega(\theta)$ amplitudes at $\theta>5$ arcsec, when the different star contamination is taken into account. Thirdly,  at $2<\theta<5$ arcsec the $\omega(\theta)$ amplitude from fields `e' is significantly 
higher  than at larger separations, by $>3.5\sigma$ at $R=22$-23 limits, whereas on field `f'
there is little difference between $A$ and $A_s$.

The  discrepancy between the two fields $A_s-A$ appears to be caused at least in part by the 
  poorer seeing on parts of field `f'. Figure 8 shows that  
at $R=21.0$--23.0 limits, the field `f' $\omega(\theta)$ falls at  $\theta<3.2$ arcsec, indicating that pairs are not being resolved, whereas on field `e' (Figure 7), this decrease occurs only at $\theta<2$ arcsec.  Hence it is
field `e' that will provide the more accurate measure of the true excess of close pairs.

\begin{table}
\caption{Estimated fraction of stars $f_s$ contaminating the galaxy samples
at each magnitude limit, and the $\omega(\theta)$ amplitudes at $\theta>5$ arcsec ($A$)
and $2<\theta<5$ arcsec ($A_s$)
multiplied by a $(1-f_s)^{-2}$ correction for star contamination.}
\begin{tabular}{lccc}
\hline
$R$ limit & $f_s$ & $A$ (corrected) & $A_s$ (corrected) \\
    & \multispan{3} \hfil Field `e'  \hfil \\
21.0 & $0.0357\pm 0.0025$ & $41.87\pm 5.35$ & $59.62\pm 6.00$ \\
21.5 & $0.1023\pm 0.0715$ & $33.50\pm 2.54$ & $49.82\pm 6.97$ \\
22.0 & $0.1217\pm 0.0851$ & $26.19\pm 1.78$ & $50.05\pm 4.40$ \\
22.5 & $0.1203\pm 0.0841$ & $22.82\pm 2.04$ & $43.04\pm 3.89$ \\
23.0 & $0.1099\pm 0.0768$ & $19.85\pm 2.36$ & $34.43\pm 3.20$ \\
23.5 & $0.0972\pm 0.0679$ & $19.72\pm 2.57$ & $29.13\pm 2.36$ \\
\smallskip
    & \multispan{3} \hfil Field `f'  \hfil \\
21.0 & $0.1342\pm 0.0066$ & $35.38\pm 3.74$ & $36.69\pm 8.97$ \\
21.5 & $0.2132\pm 0.0104$ & $24.70\pm 4.23$ & $20.95\pm 4.51$ \\
22.0 & $0.2446\pm 0.0120$ & $22.89\pm 3.26$ & $20.98\pm 4.20$ \\
22.5 & $0.2515\pm 0.0123$ & $21.42\pm 3.83$ & $24.45\pm 3.44$ \\
23.0 & $0.2447\pm 0.0120$ & $22.07\pm 3.89$ & $26.05\pm 4.30$ \\
23.5 & $0.2343\pm 0.0115$ & $23.59\pm 4.39$ & $28.31\pm 4.35$ \\

\hline
\end{tabular}
\end{table}
\onecolumn
\begin{figure}
\psfig{file=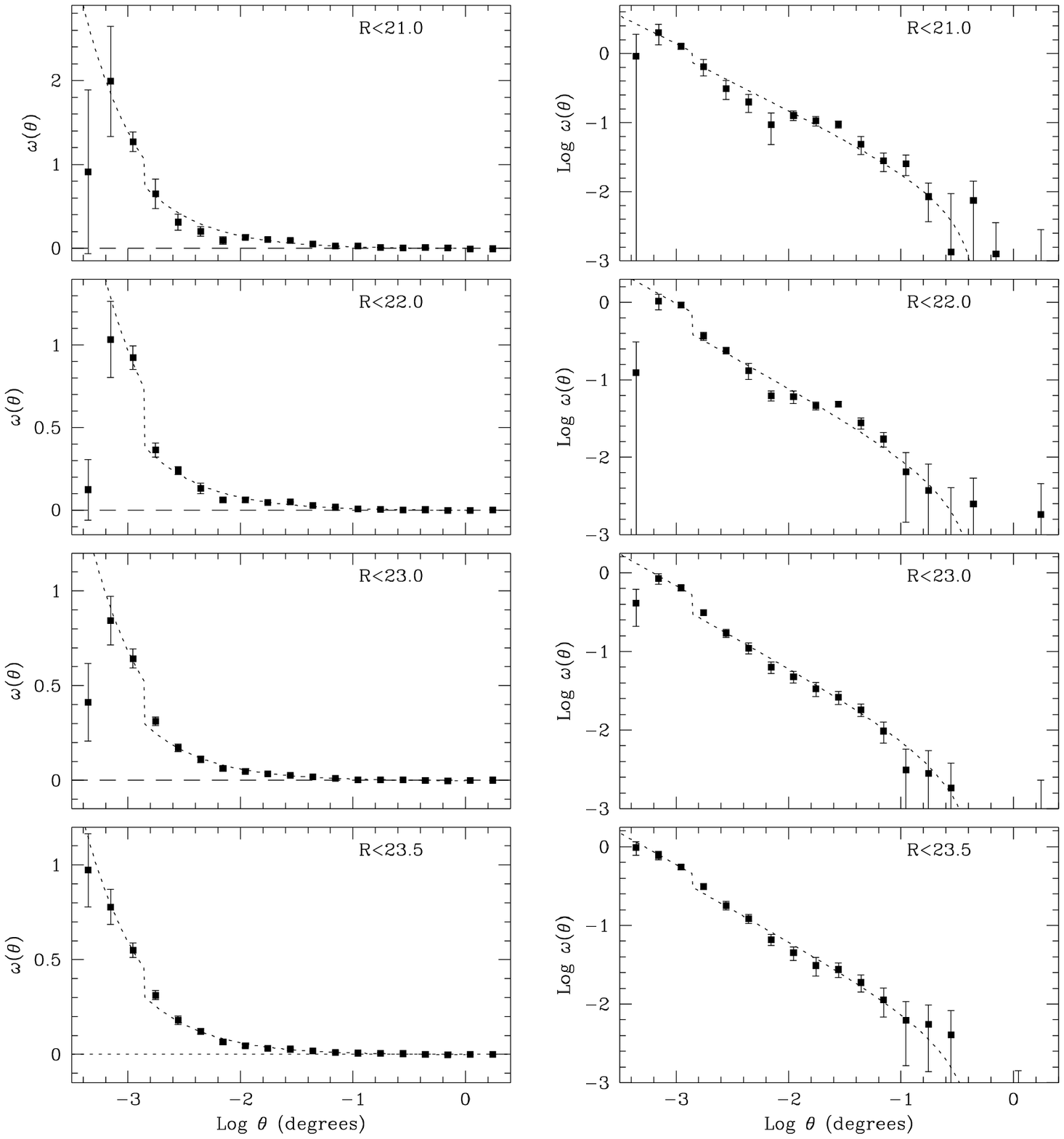,width=170mm}
\caption{Observed $\omega(\theta)$ for galaxies on field `e' brighter than 
$R=21.0$, 22.0, 23.0 and 23.5, as log-linear (left) and log-log (right) 
plots. The dotted lines show the best-fit two-part $\theta^{-0.8}$ power-laws
with the integral constraint offset.}
\end{figure}

\begin{figure}
\psfig{file=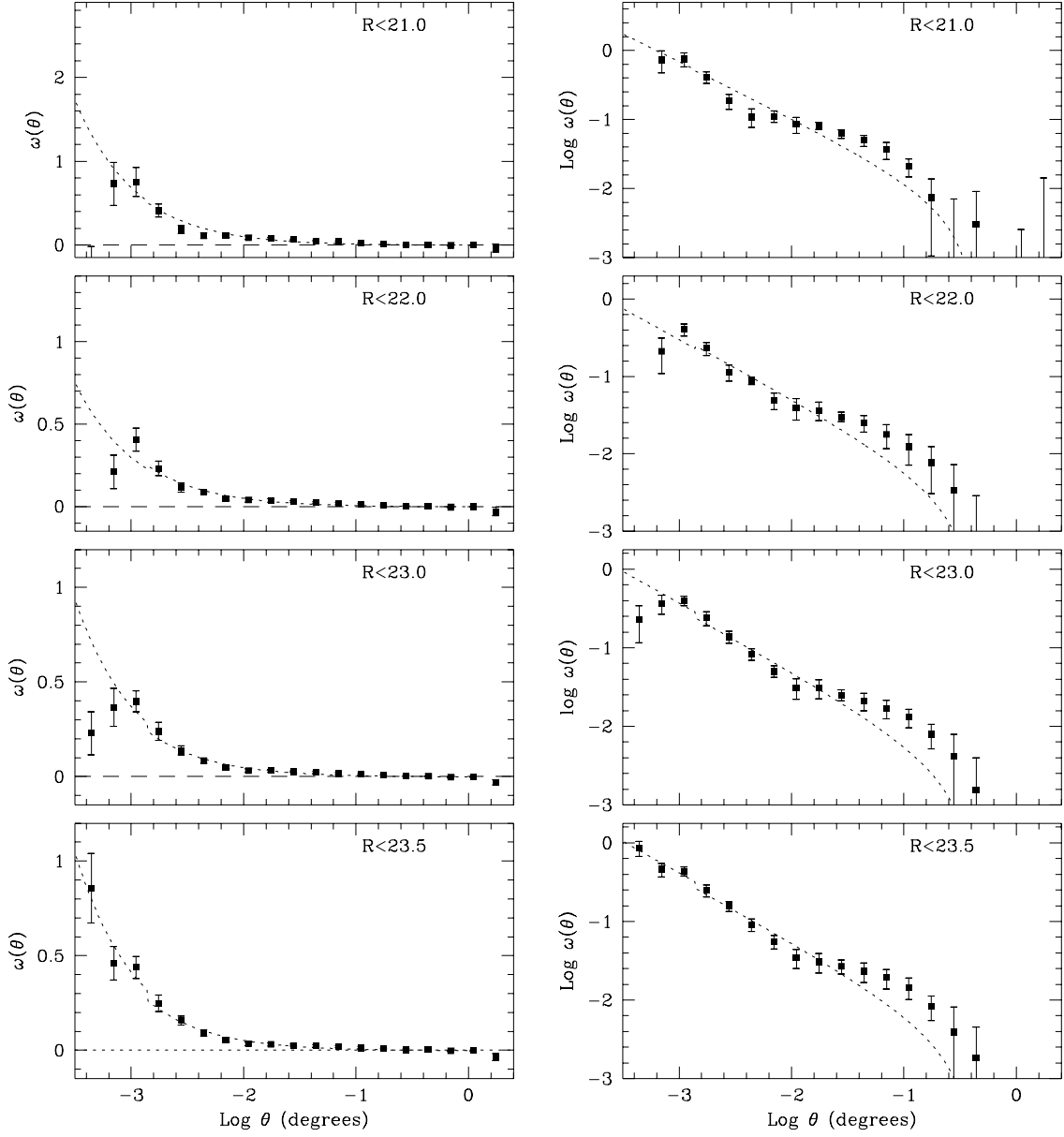,width=170mm}
\caption{Observed $\omega(\theta)$ for galaxies on field `f' fainter than $R=19.0$ and brighter than $R=21.0$, 22.0, 23.0  and 23.5, as log-linear (left) and log-log (right) 
plots. The dotted lines show the best-fit two-part $\theta^{-0.8}$ power-laws
with the integral constraint offset.} 
\end{figure}
\twocolumn

\subsection{Slope of $\omega(\theta)$}
 Infante and Pritchet (1995) found the slope of $\omega(\theta)$ from their photographic survey to flatten from $\delta=0.94\pm 0.04$  to $\delta=0.65\pm 0.05$ between limits of $R=21$ and  $R=23$, and attributed this to a flatter $\xi(r)$ for later Hubble types. However, Couch et al. (1993), using CCD data, found no significant change in the $\omega(\theta)$ slope over the same magnitude range.
We estimate a best-fit slope $\delta$ by least-squares fitting 
`$A(\theta^{-\delta}-C)$' to $\omega(\theta)$ at $\theta>5$ arcsec,
 for a range of slopes, each time recalculating the integral constraint as
\begin{equation}
C={\sum N_{rr}(\theta) \theta^{-\delta}\over \Sigma N_{rr}(\theta)}
\end{equation}

\begin{table}
\caption{Best-fitting power-laws ($\delta$) for $\omega(\theta)$}
\begin{tabular}{lcc}
\hline
$R$ limit &  Field `e' &  Field `f'  \\
     &  $\theta>5$ arcsec & $\theta>5$ arcsec \\
\smallskip
21.0 & $0.95\pm 0.14$ & $0.84\pm 0.13$ \\
21.5 & $0.90\pm 0.12$ & $0.88\pm 0.20$ \\
22.0 & $0.91\pm 0.12$ & $0.87\pm 0.18$ \\
22.5 & $0.98\pm 0.12$ & $0.87\pm 0.18$ \\
23.0 & $1.08\pm 0.15$ & $0.88\pm 0.18$ \\
23.5 & $1.07\pm 0.18$ & $0.90\pm 0.20$ \\
\hline
\end{tabular}
\end{table}
and finding the values of $\delta$ which minimize the $\chi^2$ of the fits.
Table 4 gives the best-fit $\delta$, which show no significant change over this magnitude range, on either field.
 If anything, the power-law is slightly steeper than $\theta^{-0.8}$, although this might be due to some of the effect of interactions between close pairs of galaxies extending to $\theta>5$ arcsec scales. The assumption of a $\delta=0.8$ slope in fitting an amplitude remains consistent with observations within $2\sigma$.

\subsection{Individual frame $\omega(\theta)$}
We also calculate an individual frame $\omega(\theta)$, using the methods of 
Section n except that  the 27 CCD frames of field `e' and the 20 of field `f' are treated as independent fields, with $N_{r}=20000$ randoms on each. The  
 $\omega(\theta)$ of the CCD frames within each field are averaged to give 
$\omega_{in}(\theta)$ for each of the two fields. Amplitudes $A_{in}$ were
then obtained by  fitting the
$\omega_{in}(\theta)$ at separations 5 arcsec $<\theta<21$ arcmin
with functions $A_{in}(\theta^{-0.8} - C)$, where $C=8.28$, the larger integral constraint factor calculated with equation (8) for a single CCD area.

 We find that $A_{in}$ tends to be slightly
lower than the corresponding $A$ for the full mosaiced fields (Tables 2 and 3), although the differences are less than $\sim 2\sigma$, e.g. at the $R=23.5$ limit, $A_{in}=13.58\pm 1.44\times 10^{-4}$ for field `e' and $A_{in}=9.33 \pm 2.00 \times 10^{-4}$ for field `f'  (uncorrected for star-contamination, with errors from the scatter between the $A_{in}$ of the individual CCD frames within each field) 
 
Figure 9 compares $\omega_{in}(\theta)$ with the full-field $\omega(\theta)$
at this limit.  To correct for the difference in their integral constraints, we add to $\omega_{in}(\theta)$ a positive offset of $A_{in}(C_{in}-C_{full})$, where $C_{in}$ and $C_{fr}$ are the integral constraint constants for single frames and for the full fields respectively, giving corrections of 0.00879 for field `e' and 0.00582 for field `f' at $R=23.5$. The plot also
shows the difference of the two estimates, $\Delta \omega(\theta)$.

Firstly, we note that, at small separations of $\theta<0.01$ deg, the errors on $\omega(\theta)$ derived from the scatter between the individual frame estimates are very similar to the error bars on the full-field $\omega(\theta)$ estimated by the `jackknife' method in
Section 4, as expected if the `jackknife' method is valid. 

\begin{figure}
\psfig{file=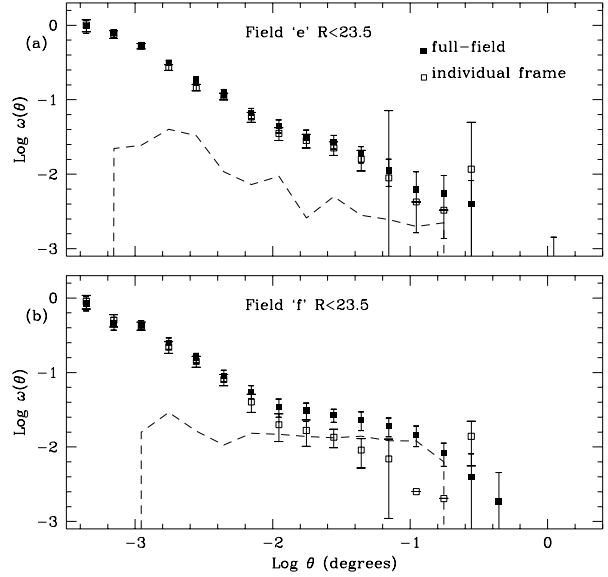,width=85mm}
\caption{Full-field  $\omega(\theta)$ and individual frame  $\omega_{in}(\theta)$, the latter with offsets added to correct for the difference in integral constraint, for $18.5\leq R\leq 23.5$ galaxies on (a) Field `e' (b) Field `f'. The
dashed lines show the difference between the two estimates $\Delta\omega(\theta)=\omega(\theta)-\omega_{in}(\theta)$, as
a function of log $\theta$.}
\end{figure}

Secondly,  $\Delta \omega(\theta)$ may give an indication of the scatter in photometric zero-points between the CCD frames forming each field. 
Although some CCD frames were corrected for photometric offsets in Section 2.5, there will inevitably be some uncorrected scatter 
$\sigma$, which will have little effect on $\omega_{in}(\theta)$ but increases the frame-to-frame variance in number counts by $(N\sigma \gamma {\rm ln }~10)^2$, where $\gamma$ is  the number counts slope (the PLE model gives $\gamma=0.39$ at $R\sim 23.5$) and $N$ the mean number of galaxies per frame. This extra variance will then increase the full-frame  $\omega(\theta)$ at $\theta$ less than the  frame size by $\Delta(\sigma)\simeq (\sigma \gamma {\rm ln}~10)^2\simeq 0.90\sigma ^2$.

However, our results suggest that, at least 
on field `e', $\Delta \omega(\theta)$
is not a constant, but contains a $\theta$-dependent component of similar slope   to $\omega(\theta)$ itself. A likely explanation  is that the 
galaxy $\omega(\theta)$ amplitude tends to be higher in areas of the data with large-scale (similar to or larger than the frame size but smaller than the field size) density
enhancements, such as rich clusters. In the individual frame $\omega(\theta)$ estimates, each frame area is given the same weighting whether or not it contains a cluster,
whereas the full-field estimate would
give higher weighting to areas with a high galaxy density. The resulting 
$\Delta \omega(\theta)$ would then contain a power-law  component in addition to a constant,  $\Delta(\sigma)$, from the photometric scatter.

To separate the two components we fit $\Delta \omega(\theta)$
with functions $A_{\Delta}\theta^{-0.8}+\Delta(\sigma)$. At $R=23.5$, this 
gives $A_{\Delta}=1.94\times 10^{-4}$ $\Delta(\sigma)=8.26\pm 9.17 \times 10^{-4}$ for field `e',  
$A_{\Delta}=1.24\times 10^{-4}$ $\Delta(\sigma)=1.095\pm 0.094 \times 10^{-2}$ for field `f'.
On both fields, $A_{\Delta}$ is about an order of magnitude less than the $\omega(\theta)$ amplitude, and therefore comparable to the $\omega(\theta)$ errors and cannot be regarded as an accurate measurement. However, a correlation between  $\omega(\theta)$ on $\theta\leq 0.1$ deg scales and galaxy density might more effectively be quantified using the moments of the galaxy counts-in-cells. As $\omega(\theta)$ is related to the variance of the galaxy counts-in-cells above the Poisson expectation, the higher variance in areas of the data with higher
counts will will result in the 
skewness and kurtosis also exceeding the Poisson expectation -- we investigate these higher moments in Section 8.

We estimate the photometric scatter $\sigma$ as $\surd {\Delta(\sigma)\over 0.90}$, giving $\sigma=0.030^{+0.014}_{-0.030}$ mag for field `e' and
 $\sigma=0.110\pm 0.005$ mag for field `f'. To estimate the likely effect on our results, we fit amplitudes to our full-field $\omega(\theta)$ with offsets subtracted 
at $\theta<0.3$ deg, of $({26\over 27})\times 8.26\times 10^{-4}$ for field `e' 
and $({19\over 20})\times 1.095\times 10^{-2}$ for field `f' (the factors in brackets allow for the partial cancelling out of the negative offset by the 
reduction in integral constraint, smaller than the offset by the ratio of the frame to the full-field area); these offsets reduced the $\omega(\theta)$ amplitude at $\theta>5$ arcsec
by 1.1 per cent on field `e' and 21.2 per cent on field `f'.
   
 Hence the effects of photometric scatter on $\omega(\theta)$ 
appear to be quite insignificant on field `e', but on field `f' are sufficient to cause a $>1\sigma$ overestimation of the amplitude at the faint limits of the data, so some sort of correction should be applied.
We fit amplitudes $A$ and $A_s$ to the field `f' $\omega(\theta)$ with the
negative offset described above, at all magnitude limits, and apply the same
star-contamination corrections as previously, giving a new set of field `f' $\omega(\theta)$ amplitudes (Table n) with corrections for both star-contamination and photometric scatter.

\begin{table}
\caption{Field `f' $\omega(\theta)$ amplitudes at $\theta>5$ arcsec ($A$)
and $2<\theta<5$ arcsec ($A_s$) with an estimated correction for the effects
of photometric scatter, in addition to the 
correction for star contamination.}
\begin{tabular}{lcc}
\hline
$R$ limit & $A$ (corrected) & $A_s$ (corrected) \\
21.0 & $30.08\pm 3.74$ & $36.14\pm 8.97$ \\
21.5 & $20.00\pm 4.23$ & $20.32\pm 4.51$ \\
22.0 & $18.31\pm 3.26$ & $20.28\pm 4.20$ \\
22.5 & $15.73\pm 3.83$ & $23.70\pm 3.44$ \\
23.0 & $16.49\pm 3.89$ & $25.33\pm 4.30$ \\
23.5 & $18.59\pm 4.39$ & $27.63\pm 4.35$ \\
\hline
\end{tabular}
\end{table}
This correction reduces the large scale $A$ much more than
$A_s$, so that some difference in these amplitudes is now seen at $R\geq 22.5$, but only at the $R=23.5$ limit does $A_s-A$ become similar to that on field `e'.
This suggetsts that photometric errors can account for part of the difference in the uncorrected  $A_s-A$ of the two fields, but the effects of poorer seeing on field `f' remain of equal or greater importance. For the remainder of the paper, we assume these new amplitudes for field `f',
but note that such corrections for the effects of photometric scatter can only be estimates  and we must regard the field `f' results as less reliable than those from field `e'.

\subsection{$\omega(\theta)$ scaling with magnitude limit} 
Figure 10 shows the scaling of the $\omega(\theta)$ amplitude with $R$ magnitude limit, for this data and several other $R$-band surveys, plotted as the amplitude at one degree when a $\theta^{-0.8}$ power-law is fitted. The magnitude limits of other surveys were converted into approximate
equivalents in our $R$ band. For the four red-band CCD surveys, we assume $R=R($Roche et al. 1993$)-0.02$, $R=R($ Roche et al. 1996$)+0.02$, $R=R($Woods and Fahlman) and $R=r($Brainerd et al.$)-0.55$, and 
 for the Infante and Pritchet (1995) $F$-band photographic survey, $R=F-0.14$ from the correction given by Metcalfe et al. (1991).
 Couch et al. (1993) used a broad `VR' passband,
and  Villumsen et al. (1997) give $\omega(\theta)$ 
amplitudes from the Hubble Deep Field as a function of $V_{606}$ limit. 
For these bluer passbands, we assume $R=VR-0.35$ and $R=V_{606}-0.39$, 
derived using the modelled spectrum of 
an evolving Sbc galaxy (Section 6.1) at $z\sim 0.5$.
 
There is some scatter between the observations, with our $\omega(\theta)$ amplitudes, especially those from field `e',  supporting the relatively high normalization of the Infante and Pritchet (1995) and the Woods and 
Fahlman (1997) $\omega(\theta)$ scaling.
The results from this data alone might suggest a levelling-out of the
$\omega(\theta)$ scaling at $R\sim 22.5$--23.5, but a comparison with  Brainerd et al. (1995), Metcalfe et al. (1995) and Villumsen et al (1997) indicates 
that, at least at these $\lambda\sim 0.4$--$0.8\mu \rm m$ passbands, the $\omega(\theta)$ amplitude falls much further beyond our survey limit.
The interpretation of these results is discussed further in Sections 6.3 and 9.

\section{Models of $\omega(\theta)$}

\subsection {Modelling of Galaxy Evolution}
In order to interpret the faint galaxy $\omega(\theta)$, we need a model which
generates a redshift distribution $N(z)$ for each type of galaxy, which is 
as consistent
as possible with observations. In this paper the 
results are compared with a Pure Luminosity Evolution (PLE)
model, i.e. a model in which the characteristic luminosity of the galaxy luminosity function, $L^*$, evolves with redshift but the normalization
$\phi^*$ and slope $\alpha$ are assumed constant.

 In the `size and luminosity evolution' model of Roche et al. (1998a),
elliptical galaxies were assumed to form in a short starburst at high redshift but spirals form by  a more gradual inside-outwards process.  Both elliptical and spiral galaxies undergo an evolutionary brightening  out to $z\sim 1$--2.
The PLE model of this paper is an updated version of this in which
(i)  the spectral evolution is now modelled using new `bc96' stellar evolutionary models (e.g. Charlot, Worthey and Bressan 1996), with a Salpeter IMF, solar metallicity for E/S0/Sab/Sbc galaxies
and 0.4 solar metallicity for Scd/Sdm and Starburst galaxies, with the same star-formation histories as previously (ii) 
dust extinction in the galaxies is now modelled as an optical depth
\begin{equation}
\tau_{dust}(\lambda)=\beta_{dust}\times {10^{10}\over M_{gal}} {dM\over dt}
\tau_{M}(\lambda)
\end{equation}
where ${dM\over dt}$ is the SFR in $M_{\odot}$ $\rm yr^{-1}$,
 $M_{gal}$ the total galaxy mass (stars and gas),
and $\tau_{M}(\lambda)$ the tabulated 
`Rv=3.1' Galactic dust extinction law of Mathis
(1990), normalized to unity at $4500\rm \AA$, and $\beta_{dust}=1.0$ for E/S0
galaxies, 0.4 for Sab spirals 0.3 for Sbc spirals and 0.12 for  Scd/Sdm/Starburst galaxies.

 \onecolumn
\begin{figure}
\psfig{file=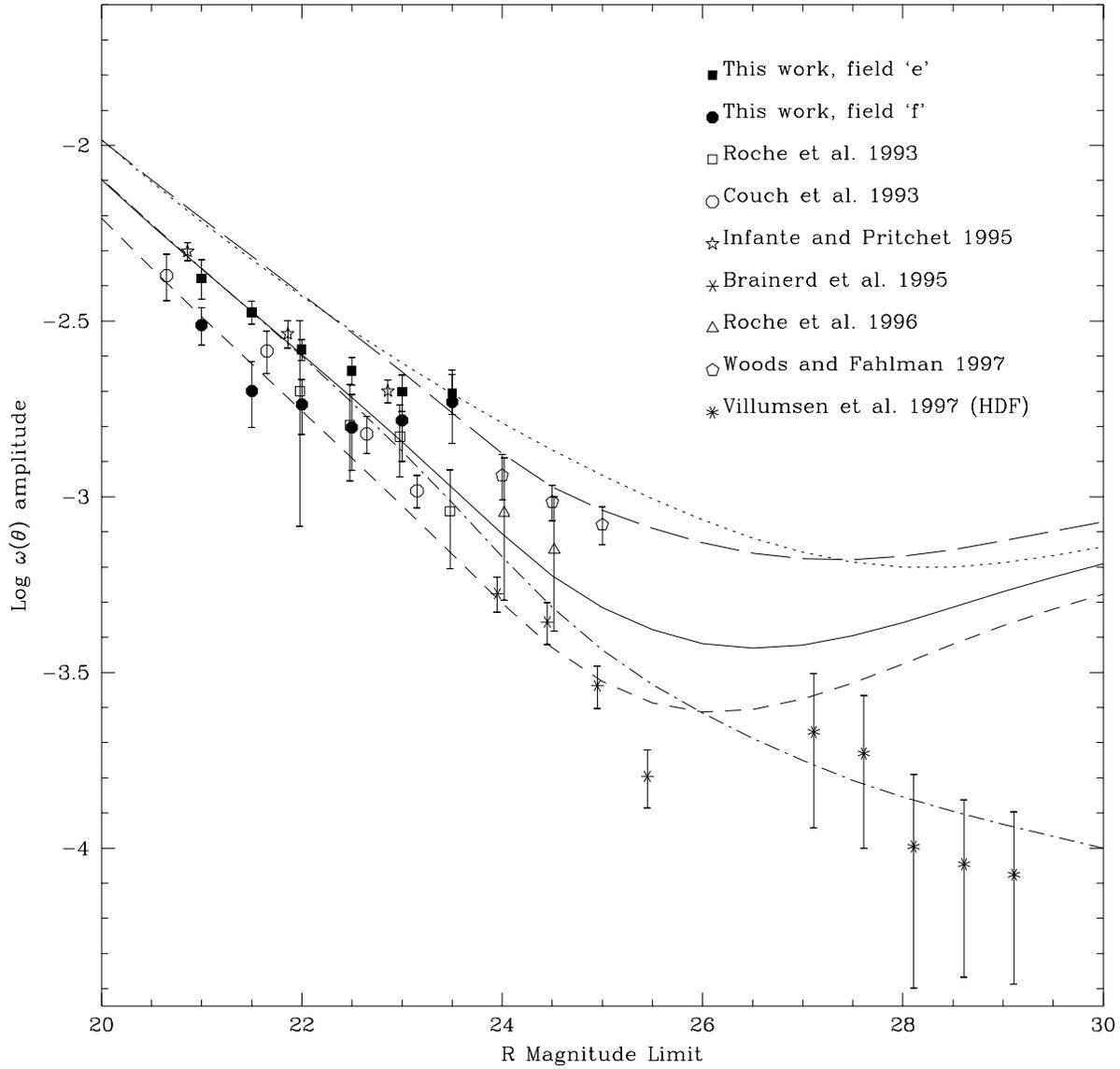,width=170mm}
\caption{The $\omega(\theta)$ amplitude at $\theta>5$ arcsec for the galaxies
in our two fields, against $R$ magnitude limit, compared with $\omega(\theta)$ amplitudes from other faint galaxy surveys and five models
described in Section 6; model A, non-evolving, $\epsilon=0$ (dotted); model B, PLE, $\epsilon=0$  (solid); model C, PLE, $\epsilon=1.2$ clustering evolution
(short-dashed); model D, PLE $\epsilon=-1.2$ comoving clustering (long-dashed); model E, PLE, $\epsilon=0$,  with a $L^{-0.25}$ decrease in the strength of clustering with luminosity (dot-dash).}
\end{figure}
\twocolumn

The galaxy luminosity functions in the model are those derived by Bromley et al. (1998) from the
red-band Las Campanas redshift survey, divided into six spectral types.
The spectral `clans' 1, 2, 3,
4 and 5 are assumed to correspond to our PLE models for E, S0, Sab, Sbc and Scd types 
respectively, while half of the galaxies in the bluest class (clan 6) are described with our evolving Scd
model (these could represent Sdm/Irr galaxies in a non-starburst phase) and the 
 other half with a 1 Gyr age starburst k-correction. We adopt the suggested 25 per cent for the fraction of light missed in this survey's 
isophotal magnitudes, asnd hence apply a correction of -0.31 mag to the
$M_R^*$, and also incorporate the estimated corrections to the luminosity
function slopes for incompleteness due to surface brightness
biases.

 Table 6 lists the  parameters of the luminosity functions, which are Schechter functions of the form $\phi(L)=\Phi^*(L/L^*)^{-\alpha}$, used in our model.

\begin{table}
\caption{Luminosity function normalizations $\Phi^*$ (in galaxies 
$\rm Mpc^{-3}$ for $H_0=50$ km $\rm s^{-1}Mpc^{-1}$, slopes $\alpha$ and 
characteristic $R$-band absolute magnitudes $M^*$ (for  $H_0=50$ km $\rm s^{-1}Mpc^{-1}$)}
\begin{tabular}{lccc}
\hline
Galaxy type & $\Phi^*$ & $\alpha$ &  $M^*_R$  \\
\smallskip
E   & $0.48\times 10^{-4}$ & +0.54 & -22.14 \\
S0  & $8.6125\times 10^{-4}$  & -0.12 & -22.09 \\
Sab & $10.0675\times 10^{-4}$  & -0.32 & -21.76 \\
Sbc & $9.1625\times 10^{-4}$  & -0.71 & -21.71 \\
Scd & $2.125\times 10^{-4} $ & -1.45 & -21.89 \\
Sdm & $1.10625\times 10^{-4}$ & -1.89 & -21.87 \\
\hline
\end{tabular}
\end{table}

For comparison, we also consider a non-evolving model, with the same 
$z=0$ luminosity functions, and k-corrections derived by redshifting without evolution the
galaxy spectra of the PLE models at $z=0$.

 The $L^*$ evolution in the PLE model increases the predicted
galaxy number counts sufficiently to fit the observations(Figure 5).
Figure 11 shows the galaxy redshift distributions $N(z)$ predicted by both 
models at $R=21$--23.5 limits.
In the PLE model,  a `tail' of $1<z<2$ galaxies appears between 
$R=22$ and $R=23$, whereas with no $L^*$ evolution very few $z>1$ galaxies
would be seen even at the survey limit of $R=23.5$.  
Deep spectroscopy (Cowie et al. 1996)
reveals that $1<z<1.8$ galaxies do start to appear in substantial numbers at $B>23.2$ ($R\ga 22.7$), in general agreement
with this PLE model. The PLE model is also reasonably consistent with the
numbers and surface brightness of the $z>2.5$ `Lyman break' galaxies on the 
Hubble Deep Field, although these more distant galaxies are not expected to be visible within this survey.

Figure 12a shows the mean redshift of both models as a function of $R$ magnitude limit. At
the survey limit of $R=23.5$, the PLE model predicts $z_{mean}=0.685$, which
 increases on going further faintward to $z_{mean}\simeq 1.4$ at $R\sim 27$. 
As our PLE model seems to fit the galaxy counts and $N(z)$ reasonably well, we now use it in interpreting the $\omega(\theta)$
amplitudes.
\begin{figure}
\psfig{file=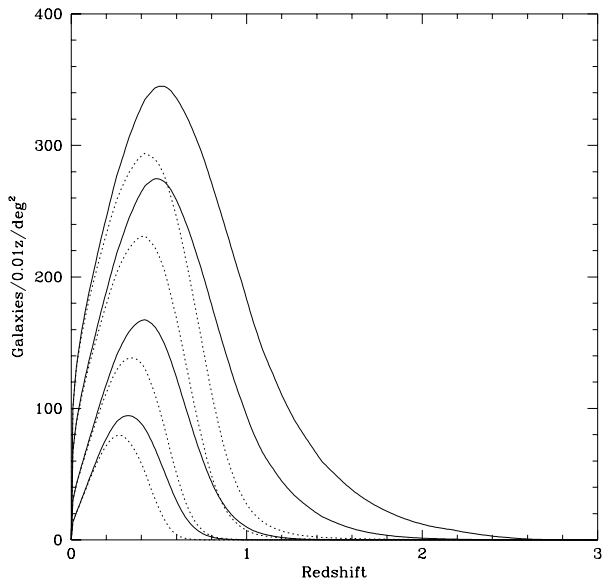,width=85mm}
\caption{Galaxy redshift distributions predicted for all galaxies to limits of
(from lowest to highest), 
$R=21.0$, 22.0, 23.0 and 23.5, for the PLE (solid) and no-evolution (dotted) models.}
\end{figure}

\begin{figure}
\psfig{file=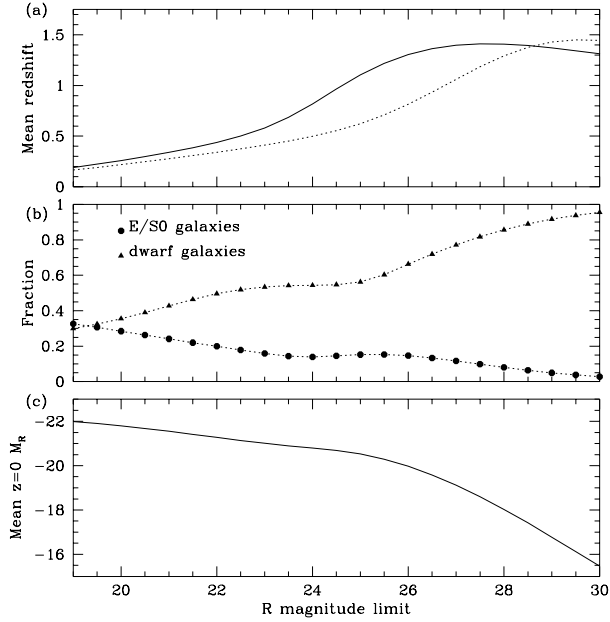,width=85mm}
\caption{(a) Mean redshift as a function of $R$ magnitude limit for PLE (solid)
and non-evolving (dotted) models. (b) The fractions of E/S0 galaxies and dwarf galaxies (those with $M_B>-20.5$ at $z=0$) as  a function of $R$ magnitude limit for the PLE model. (c) The mean $M_R$ at $z=0$ as a function
of magnitude limit in the PLE model.}
\end{figure}

\subsection{From $N(z)$ to $\omega(\theta)$}

The $\omega(\theta)$ amplitude is predicted
using the Limber's formula integration of $\xi(r,z)$ over the modelled 
$N(z)$. We assume the form given in the Introduction,
\begin{equation}   
\xi(r,z)=(r/r_0)^{-\gamma}(1+z)^{-(3+\epsilon)}
\end{equation}
 where $r_0$ normalizes the strength of clustering
at $z=0$, $\gamma$ is the slope and $\epsilon$ parameterizes the clustering 
evolution relative to the $\epsilon=0$ stable clustering model.

The $\omega(\theta)$ amplitude will generally decrease on going faintward, 
primarily as a result of $N(z)$ becoming more extended, but there may
be additional effects due to the evolution of clustering and to changes in the 
proportions of different types of galaxy.
 The model of 
Roche et al. (1996) assumed $\gamma=1.8$ for all galaxy types, with
$r_0=5.9$ $h^{-1}$ Mpc for E/S0 galaxies
and $r_0=4.4$ $h^{-1}$ Mpc for later types
 (from Loveday et al. 1995, hereafter LMEP) and
also dwarf galaxies (those with $z=0$ blue-band absolute magnitudes $M_B>-20.5$)  half as clustered as 
those of higher luminosity (again from LMEP).

With stable clustering and $L^*$ evolution, this model predicted a $\omega(\theta)$ scaling fairly  consistent with observations in the blue-band (Roche et al. 1996), but it underpredicted the 
$\omega(\theta)$ amplitudes from $K$-band surveys at $K\simeq 19.5$--21.5 limits (Roche et al. 1998b; Carlberg et al. 1997). The results from the $K$-band surveys, in which E/S0 galaxies will be more prominent, were fitted better by 
assuming even stronger clustering for early-type galaxies (e.g. as observed by Guzzo et al. 1997). 

The $r_0$ used by Roche et al. (1996) were derived by LMEP from a 
cross-correlation of each type of galaxy with all APM galaxies, a
method which would
tend to underestimate the differences between the $\xi(r)$ of the galaxy
types. LMEP also derive $\xi(r)$ using a different `inversion' method, which gave $r_0=7.76\pm 0.15$ $h^{-1}$ Mpc and
$\gamma=1.87\pm 0.07$ for E/S0 (of all luminosities) and 
$r_0=4.49\pm 0.13$ $h^{-1}$ Mpc and $\gamma=1.72\pm 0.05$ for late-types. These estimates would be more consistent with Guzzo et al. (1997) and the results from the $K$-band.
In this paper we model $\xi(r)$ with, for simplicity, a single slope $\gamma=1.8$, and take the normalization  from the LMEP `inversion' estimates at $r=1 h^{-1}$ Mpc. This gives for E/S0 galaxies
 $(r_0)^{1.8}=7.76^{1.87}$ and thus $r_0=8.4$ $h^{-1}$ Mpc,
and for later types $(r_0)^{1.8}=4.49^{1.72}$ giving $r_0=4.2$ $h^{-1}$ Mpc.

In the Limber's formula integration, 
we normalize our models to  the LMEP spiral
$r_0$ and then apply a  weighting term to the galaxy clustering at 
each redshift  derived
from the predicted fraction of ellipticals ($f_e$). With our assumed  $r_0$, $\xi_{elliptical}\simeq 3.5 \xi_{spiral}$, and if the cross-correlation between
early and late type is
assumed to be the geometric mean of their autocorrelations so that 
  $\xi_{cross}\simeq 1.87\xi_{spiral}$, the adopted weighting factor is
$(1-f_e)^{2}+3.5f_e^{2}+2\times
1.87f_e(1-f_e)$ -- or equivalently $(1+0.87f_e)^2$.

The luminosity dependence of clustering is modelled by two methods. The first   is as in Roche et al. (1996) --
according to LMEP, lower luminosity ($M_B>-20.5$) galaxies are less clustered than more luminous galaxies of the same Hubble type by a factor of 
2 at $r\sim 1$ $\rm h^{-1} Mpc$, so, at each redshift, the models calculate $f_d$, the fraction of galaxies which would
have $M_B>-20.5$ at $z=0$, and apply a luminosity weighting, 
normalized to be unity for the dwarf fraction in the LMEP
dataset, of $1.2(1-f_d)^{2}+0.6f_d^{2} +2\times 0.85(1-f_d)f_d$.

Figure 12b shows $f_{e}$ and $f_{d}$ as a function of $R$ limit in the PLE model -- $f_{e}$ slowly decreases (from 0.28 to 0.18 between
$R\leq 21$ and $R\leq 23.5$) while $f_{d}$ increases.  Both these
trends would slightly steepen the scaling of $\omega(\theta)$.
Dwarf galaxies (almost all of late type, at least in the
field environment) will dominate the sample at $R\ga 26$, as $L^*$
galaxies are seen to their highest redshifts at these magnitudes and going fainter simply results in looking further down the faint-end of the galaxy luminosity function for the same volume of space.

Using the luminosity evolution and $r_0$ described above, we predict the
scaling of $\omega(\theta)$ with $R$ magnitude limit for four models

\noindent A: $N(z)$ from non-evolving model, $\epsilon=0$

\noindent B: $N(z)$ from PLE model, $\epsilon=0$

\noindent C: $N(z)$ from PLE model, clustering evolution of $\epsilon=1.2$. 

\noindent D: $N(z)$ from PLE model, comoving clustering of $\epsilon=-1.2$.

We also consider a fifth model E, which differs from model B only in the luminosity dependence of $\xi(r)$. Here the strength of clustering varies continuously with the unevolved luminosity (i.e. very approximately the mass) of the galaxies, as $\xi(r)\propto L^{0.25}$. 
The luminosity weighting term in Limber's formula is then 
$10^{-0.1(\langle M_R \rangle +22.0)}$, where $\langle M_R \rangle$ is the mean
unevolved (i.e. at $z=0$) $R$-band absolute magnitude of all galaxies at a given redshift,
 and the addition of 22.0 normalizes the model to the same $\omega(\theta)$ amplitude as
model B at $R\sim 22.0$.
Model E would be consistent with the dependence of $\xi(r)$ on absolute magnitude as observed (LMEP, Guzzo et al. 1997) for relatively luminous galaxies,  but includes the additional assumption that clustering  
continues to decrease below $M_B=-20$, becoming very weak for the faintest dwarfs.  Models B and E would be indistinguishable at the limits of our survey
but differ greatly at $R\ga 26$ where the 
mean unevolved luminosity (Figure 12c), which decreases only slowly to  $R\sim 25.5$, falls more steeply as the sample becomes dominated by dwarf galaxies.

\subsection{Comparison with Observations}

Figure 10 shows the five models against the $\omega(\theta)$ observations.
Model A gives $\omega(\theta)$ amplitudes well above most of the data and can be
rejected (redshift
surveys have already excluded models with no $L^*$ evolution). To $R\sim 25$, PLE model B passes through the middle of the rather scattered 
data points and is the most consistent with our results; however it 
overpredicts the very weak clustering of the faintest Hubble Deep Field galaxies.

Model C, with clustering evolution of $\epsilon=1.2$, passes through only the lowest of the data points at $R<25$. At $R=21$--23.5, $\epsilon=1.2$ clustering evolution
reduces $\omega(\theta)$ 0.134--0.191 dex below the model B prediction. It 
significantly underpredicts our amplitudes for field `e' at all limits and for field `f' at $R=23$--23.5, and appears to be 
 rejected by $\geq 3\sigma$. The comoving model D gives a very similar $\omega(\theta)$ 
scaling to the non-evolving model A, overpredicting the faint galaxy clustering, and can be rejected.

If we assume the $N(z)$ and $r_0$ in our PLE model to be correct and treat
$\epsilon$ as a free parameter, a comparison of the model at $R=21$--23.5 with 
the combined field `e' and field `f' results
minimizes $\chi^2$ for $\epsilon= 0.02_{-0.31}^{+0.48}$. The same comparison for the results from the more photometrically reliable and less star-contaminated field `e' 
only gives a consistent  $\epsilon= -0.29_{-0.27}^{+0.31}$, and rejects 
$\epsilon\geq 0.77$ and $\epsilon\leq -1.03$ by $\geq 3\sigma$.   
 
To $R\sim 24$, the model E $\omega(\theta)$ does not differ significantly from model B, so it is similarly consistent with our results, but at $R>26$ the decrease in $r_0$ with luminosity causes the $\omega(\theta)$ amplitude to continue falling after $z_{mean}$ has reached its maximum, in contrast to  models A-D where the increasing number of low redshift dwarfs causes $\omega(\theta)$ to start increasing again.
 Model E, in contrast to the other four, fits the Hubble Deep Field clustering very well.

\section{The Small-scale $\omega(\theta)$ and Merger Rate Evolution}

  Infante et al. (1996) found that the  $\omega(\theta)$ of a sample of $R<21.5$ galaxies,
remained a $\delta\simeq 0.8$ power-law at $\theta<6$ arcsec but with an amplitude a factor
1.8 higher than at $\theta>6$ arcsec. We investigated this effect in our data (Section 4) by fitting two separate $\omega(\theta)$ amplitudes $A$ and $A_s$ at $\theta>5$ arcsec and  $2<\theta<5$ arcsec respectively, and found that on field `e' $A_s$ was significantly higher than $A$, at all magnitude limits. 
The results from field `f' are not used in the following analysis, as although the
corrected amplitudes for this field did show some excess of $A_s$ over $A$ at the faintest limits (Table n), any estimate of $A_s-A$ will be very dependent on
our uncertain correction for photometric scatter, and even after this correction $A_s$ would be, at least at $R\leq 23$, affected by the poorer seeing.  

At the mean redshift of the Infante et al. (1996) sample, $z_{mean}\simeq 0.35$,
$\theta\leq 6$ arcsec corresponds to a projected physical separation
 $\la 20$ $h^{-1}$ kpc. Even locally, there is an excess of $\la 20$ $h^{-1}$ kpc pairs above the number expected from normal galaxy clustering, made  up of interacting galaxies including those in the early stages of merging. In a magnitude-limited ($B<14.5$) $z_{mean}=0.007$ subsample of the UGC catalog, $4.6\pm 0.4$ per cent of galaxies
are in physical pairs of projected separation $<19$ $h^{-1}$ kpc (Carlberg et al. 1994). 
We derive a similar `close pair fraction', $f_{pairs}$, for our field `e', here defining $f_{pairs}$ as the fraction of the galaxies which are in $\theta<5$ arcsec
pairs, above the expectation from the $\omega(\theta)$ measured at $\theta>5$ arcsec.

 Firstly, we assume that $\omega(\theta)\propto \theta^{-0.8}$ at
all $\theta<5$ arcsec, so that we can estimate the number
of $\theta<2$ arcsec pairs from an inwards extrapolation of this power-law with its $2<\theta<5$ arcsec amplitude. In an unbounded area, with a surface density of galaxies $\rho_{gal}$, the number of
$\theta<\beta$ pairs per galaxy above the random expectation will be
\begin{equation}
{N_{pairs}\over N_{gal}}=\rho_{gal}A\int_{0}^{\beta}2\pi\theta\theta^{-0.8} d\theta
\end{equation}
where $A$ is the $\omega(\theta)$ amplitude. To find the number of pairs above
that expected from the $\omega(\theta)$ seen at 
larger scales, $A$ is replaced by the difference in clustering amplitudes
$A_s-A$, so that
\begin{equation}
{N_{pairs}\over N_{gal}}=\rho_{gal}(A_s-A)
\int_{0}^{\beta}2\pi\theta\theta^{-0.8} d\theta
\end{equation}
which for $\beta=5$ arcsec gives
\begin{equation}
{N_{pairs}\over N_{gal}}=0.00196 \rho_{gal} (A_s-A)
\end{equation}
In a finite field $N_{pairs}$ will be reduced as some galaxies will lie less
than 5 arcsec from the field edges or the holed areas, but this effect is minor for the large area of our fields -- a numerical integration over the field `e' area gives
 \begin{equation}
{N_{pairs}\over N_{gal}}=0.001938 \rho_{gal} (A_s-A)
\end{equation}
As a pair contains two galaxies, $f_{pair}=2{N_{pairs}\over N_{gal}}$.
We estimate $f_{pairs}$ at each magnitude limit using the field `e' $A$ and $A_s$ corrected for
star contamination (Table 3), and also apply a star contamination correction to the observed $\rho_{gal}$,
 \begin{equation}
f_{pairs}=0.003876 (1-f_{star})\rho_{gal} (A_s-A)
\end{equation}
which is given in Table 7, with errors from adding the
errors on $A$ and $A_s$ in quadrature. The excess of close pairs above the $\omega(\theta)$ measured at larger scales is $3.7\sigma$ significance at $R\leq 22.5$ and $>3\sigma$ at $R\leq 22$ and $R\leq 23$. 
\begin{table}
\caption{Percentage of galaxies in excess $\theta<5$ arcsec pairs on field
`e'}
\begin{tabular}{lc}
\hline
$R$ limit & $f_{pairs}$ (percentage) \\
21.0 & $4.03\pm 1.85$ \\
21.5 & $5.14\pm 2.08$ \\
22.0 & $8.32\pm 2.60$ \\
22.5 & $12.33\pm 3.36$ \\
23.0 & $12.88\pm 4.05$ \\
23.5 & $11.12\pm 4.86$ \\
\hline
\end{tabular}
\end{table}
    
Infante et
al (1996) found 1317 pairs with $2<\theta<6$ arcsec in a sample of 16749 galaxies with $R<21.5$, compared to 477 expected by chance or 842
expected from the clustering at larger scales. The proportion of galaxies
in $2<\theta<6$ pairs above the large-scale $\omega(\theta)$ was
then $2\times {1317-842 \over 16749}=5.67$ per cent. For a $\theta^{-0.8}$ power-law
the number of $\theta<5 $ arcsec pairs will be 1.097 times the number at
 $2<\theta<6$ arcsec, and the significance of the pairs excess is given as 
$\sim 5\sigma$, so the Infante et al. (1996)
result corresponds to $f_{pair}=6.22\pm 1.24$ per cent.

As the timescale between the approach of two galaxies within $\sim 20$ $h^{-1}$ kpc and their subsequent merging or separation, $\sim 0.75$ Gyr
(e.g. Mihos and Hernquist 1996), is short compared to  the Hubble time, the merger/interaction rate $R_{m}\propto f_{pair}$ approximately. The merger rate evolution 
is modelled as $R_{m}\propto
R_{m0}(1+z)^m$ and normalized to a 4.6 per cent fraction of local galaxies in pairs of projected separation $<19$ $h^{-1}$ kpc (Carlberg et al. 1994).
Pairs within this separation will have an angular separation $\theta<5$ arcsec at angular diameter distance $d_A>784$ $h^{-1}$ Mpc ($z>0.455$ in our
chosen cosmology). At $d_A<784$ $h^{-1}$ Mpc, if we again assume  
$\omega(\theta)\propto \theta^{-0.8}$ 
for the close-pair galaxies, the fraction with $\theta<5$ arcsec
will be $({d_A(z)\over 784 h^{-1}})^{1.2}$.

For each magnitude limit we then model $f_{pair}$ by summing
\begin{equation}
f_{pair}=f_{pair}(z<0.455)+f_{pair}(z>0.455)
\end{equation}
where
\[
f_{pair}(z<0.455)=0.046 {\int_{0}^{0.455}({d_A(z)\over 784 h^{-1}})^{1.2} N(z)(1+z)^m dz  \over \int_{0}^{0.455}N(z)dz}
\]
\[
f_{pair}(z>0.455)=0.046 {\int_{0.455}^{6}N(z)(1+z)^m dz  \over \int_{0.455}^{6}N(z)dz}
\]
over the PLE model $N(z)$.

 Infante et al. (1996) estimated $m=2.2\pm 0.5$ from the
 increase in the fraction of galaxies in close pairs from $z=0$ to
 $z_{mean}=0.35$.
 Our data suggest that the upward trend in $f_{pair}$ continues to fainter  magnitudes,
with the slight drop from $R=23$ to $R=23.5$ probably being due to less
effective splitting of close pairs on approaching the limits of the data 
($f_{pair}$ decreases steeply beyond the $R=23.5$ limit). 
\begin{figure}
\psfig{file=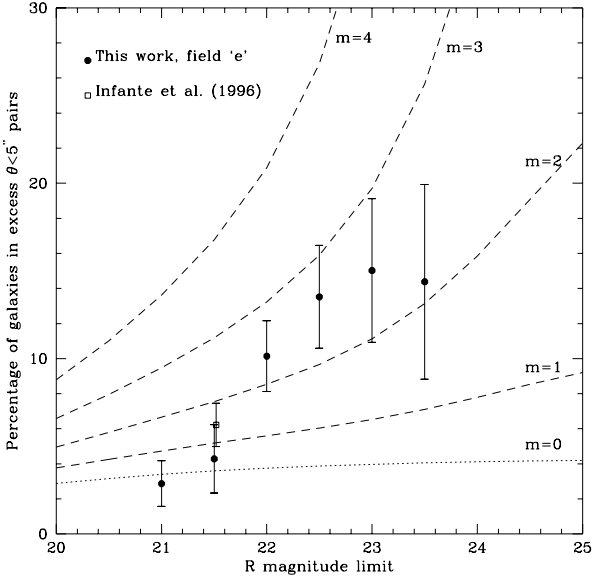,width=85mm}
\caption{Fraction of galaxies in $\theta<5$ arcsec pairs in excess of the
number expected by chance and the clustering observed at $\theta>5$ arcsec, as a
function of $R$ magnitude limit, from our field `e' and from  Infante et al. (1996). The dotted line shows a model in which the local fraction of galaxies in close pairs remains constant, the dashed lines are models in which the merger rate evolves as $(1+z)^m$, where $m=1,2,3,4$.}. 
\end{figure}

Figure 13 shows $f_{pair}$ from our field `e' and from Infante et al. (1996), compared with the model (equation 18) for $m=0,1,2,3$ and 4. The $m=2$ model is the most consistent with the data; both a non-evolving merging rate ($m=0$) and very rapid evolution of $m=4$ appear
to be rejected. Our $R=21.0$ point lies significantly below the $m=2$ model, but
at the brighter limits there might be some underestimation of $f_{pair}$ if the larger angular sizes of brighter galaxies prevents the resolution of some $2<\theta<5$ arcsec pairs. A $\chi^2$ test of our model against the plotted data (excluding
the $R=21.0$ point) gives a best-fitting merger rate evolution of $m=2.01_{-0.69}^{+0.52}$, with evolution of $m\geq 3.25$ rejected by 
$\geq 3\sigma$. This is discussed further in Section 9.2.

\section{Higher-order Correlation Functions}

The large size of our galaxy sample enables us to investigate their 
higher order correlation functions, which may discriminate between models of galaxy clustering more effectively than $\omega(\theta)$ alone. The higher order correlations have previously been studied at brighter magnitudes of $17\leq B\leq 20$, for $1.6\times 10^6$
 galaxies in the APM survey (Gazta\~{n}aga 1994) and $2.9\times 10^5$ galaxies 
Edinburgh-Durham Southern Galaxy Catalog Survey (EDSGCS) (Szapudi, Meiksin and
Nichol 1996), in the form of hierarchical moments.
 Before discussing these results further, we define the hierarchical moments
and describe their estimation from a simple counts-in-cells method.
\subsection{Estimation of Hierarchical Moments}
Consider a survey area divided into $m$ cells of area
$\Omega$, with $N_i$ galaxies in cell $i$ and a mean number of galaxies per cell $\bar{N}$. The moments of the distribution of counts-in-cells will be 
\begin{equation}
\mu_J={1\over m}\sum^{m}_{i=1}(N_i-\bar{N})^J.
\end{equation}
and the second moment (variance) of the
counts-in-cells will be related to $\omega(\theta)$ as
\begin{equation}
\mu_2=\bar{N}+\bar{N}^2\int\int_{cell} \omega(\theta) d\Omega_1 d\Omega_2
\end{equation}
where the double integral is over the area of one cell.
An area-averaged $\omega(\theta)$, $\bar{\omega}_2$, can be defined as 
\begin{equation}
\bar{\omega}_2=\int\int \omega(\theta) d\Omega_1 d\Omega_2
\end{equation}
and estimated from the measured second moment by subtracting the expectation
for a Poission distribution (i.e. $\mu_2=\bar{N}$) and dividing by $\bar{N}^2$.
\begin{equation}
\bar{\omega}_2={1\over \bar{N}^2}(\mu_2-\bar{N})
\end {equation}
Similarly, the area-averaged angular three-point correlation function $\bar{\omega}_3
$ can be calculated from the excess of the third moment (skewness) relative to the Poission expectation (Baugh, Gazta\~{n}aga and 
Efstathiou 1995)
\begin{equation}
\bar{\omega}_3={1\over \bar{N}^3}(\mu_3-3\mu_2+2\bar{N})
\end {equation}
and the area-averaged angular four-point correlation function $\bar{\omega}_4$ can be
 calculated from the excess of the fourth moment (kurtosis) relative
to the Poission expectation
\begin{equation}
\bar{\omega}_4={1\over \bar{N}^4}(\mu_4-3\mu_2^2+11\mu_2+6\bar{N})
\end {equation}
It is useful to express higher order correlation functions in the form
of hierarchical amplitudes $s_J$, defined as
\begin{equation}
s_J={\bar{\omega}_J\over \bar{\omega_2}^{J-1}}
\end{equation}

We divide fields `e' and `f' into square cells of side $\theta$, and count the number
 of  $18.5\leq R\leq 23.5$ galaxies in each cell.
 Many of the cells will overlap with field edges or holed areas. If the missing proportion of the cell area $f_{m}$ is less than 0.3,
the galaxy count is corrected by multiplying by $(1-f_{m})^{-1}$, while
cells with $f_{m}>0.3$ were excluded from the analysis (as in Gazta\~{n}aga
1994). Combining fields `e' and `f', we calculate the variance, skewness and kurtosis of the counts-in-cells, for cell size $\theta$ ranging from 0.01 to 0.1 degrees.

As for $\omega(\theta)$, 
$\bar{\omega}_2(\theta)$ is corrected for star-contamination by multiplying by $(1-f_{s})^{-2}$,
where the estimated star-contamination $f_{s}=0.15866\pm 0.00634$
for fields `e' and `f' combined.
Error bars were 
calculated by randomly dividing the cells into 10 subsamples, calculating
 $\bar{\omega}_2(\theta)$ for each subsample the scatter between them
by $\surd 10$, finally adding in quadrature the errors from 
the uncertainty
in the star-contamination correction.
 
\begin{figure}
\psfig{file=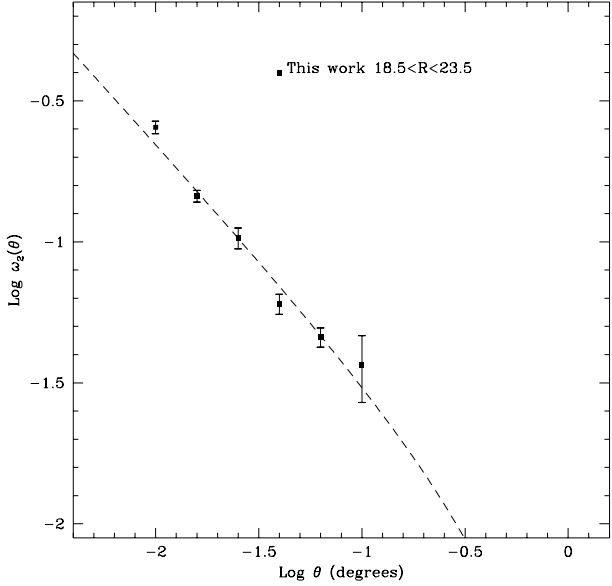,width=85mm}
\caption{The area-averaged angular correlation function 
$\bar{\omega}_2(\theta)$, from the counts-in-cells analysis of both fields combined, with the best-fitting function $A(\theta^{-0.8}-0.9606)$.}
\end{figure}

As for $\omega(\theta)$, an amplitude is estimated by fitting with a
function $A(\theta^{-0.8}-C)$, where the integral constraint factor $C$ is derived from those of the two independent fields, $C_e$ and $C_f$ (see Section n) and their areas $\Omega_e$ and $\Omega_f$, as
\begin{equation}
C={\Omega_e^2C_e+\Omega_f^2C_f\over (\Omega_e+\Omega_f)^2}
\end{equation}
giving $C=0.9606$.
Figure 13 shows  $\bar{\omega}_2(\theta)$ with the best-fitting function  
$A(\theta^{-0.8}-0.9606)$, with $A=56.8\pm1.3\times 10^{-4}$ at a cell size one degree. As the integral constraint factor $C=2.24$ for a $1\times1$ deg cell, this corresponds to a $\omega(\theta)$ amplitude of ${A\over 2.24}=25.4\pm 0.6\times 10^{-4}$, slightly higher than 
but reasonably consistent with the amplitudes calculated from the positions 
of individual galaxies.  
 
To correct the hierarchical moments for the integral constraint in 
 $\bar{\omega}_2(\theta)$, estimated as $AC=0.00387$ from the fitted 
 $\bar{\omega}_2(\theta$ amplitude before correction for star-contamination, $s_3(\theta)$ and $s_4(\theta)$
 are calculated as 
\begin{equation}
s_J(\theta)={\bar{\omega}_J(\theta)\over(\bar{\omega_2(\theta)}+0.00387)^{J-1}}
\end{equation}
No corrections are applied for the integral constraints in  
$\bar{\omega}_3(\theta)$
and  $\bar{\omega}_4(\theta)$, but these will be much smaller than for  
$\bar{\omega}_2(\theta)$, due to the much steeper
slopes of the higher-order correlation functions.

Star contamination
will reduce $\bar{\omega}_3(\theta)$ by a factor $(1-f_{s})^{-3}$ but reduces the denominator
$\bar{\omega}_2(\theta)^2$ by $(1-f_{s})^{-4}$, so $s_3(\theta)$ is actually overestimated, and can be corrected for star-contamination by multiplying
by $(1-f_{sr})$. Similarly, star contamination
will reduce $\bar{\omega}_4(\theta)$ by $(1-f_{s})^{-4}$
 but reduces $\bar{\omega}_2(\theta)^3$ by $(1-f_{s})^{-6}$, so $s_4(\theta)$
is corrected by multiplying by $(1-f_{s})^2$.

By a similar argument, frame-to-frame photometric scatter increases $\bar{\omega}_2(\theta)$ over this range of cell sizes, and so is expected to decrease
$s_3(\theta)$ and $s_4(\theta)$, with the effect on  $s_3(\theta)$ being about half that for 
$\omega(\theta)$. The photometric scatter on field `f' was estimated to reduce
the field's $\omega(\theta)$ amplitude, fitted over a similar range of $\theta$, by 21.2 per cent (Section 5.3), so for both fields combined its effect on 
$s_3(\theta)$ should be $\leq 5$ per cent, much less than $1\sigma$. However, it is possible that photometric or other differences between the two fields could also affect
results for the combined fields. In order to check the
validity of using the combined fields, we compare the results with $s_3(\theta)$ and $s_4(\theta)$ derived by performing the same counts-in-cells analysis for the cells in field `e' only, applying corrections for star-contamination and integral constraint appropriate for this field alone.
\subsection{Results for $s_3(\theta)$ and $s_4(\theta)$}
Figures 14 and 15 show $s_3(\theta)$ and  $s_4(\theta)$, corrected for both 
star-contamination and the integral constraint. The $s_3(\theta)$ and  $s_4(\theta)$ of the field `e' data only are very similar
in both slope and normalization to those from the full dataset, supporting the 
accuracy of the combined field result.
There is a highly significant
($\sim 8\sigma$) detection of a positive signal in $s_3(\theta)$, and it is
clear that $s_3(\theta)$ is not constant with increasing $\theta$ but decreases quite steeply over this range of cell size -- the best-fitting $\theta^{-\delta}$ power-law is of slope $\delta=0.413\pm 0.083$. The statistics are poorer for $s_{4}(\theta)$ but the
detection is still 4--$5\sigma$ and the best-fit slope of $\delta=0.614\pm 0.134$ is even steeper.

We compare our results with the $s_3(\theta)$ and $s_4(\theta)$ of
$17\leq B\leq 20$ galaxies in the 
APM (Gazta\~{n}aga 1994) and EDSGCS (Szapudi et al. 1996) surveys.
The hierarchical moments are projections of the corresponding three-dimensional
$S_J(r)$, defined as the ratios of the three-dimensional J-point correlation functions to $\xi(r)^{J-1}$.
Hierarchical moments are well-suited for a comparison of deep and shallower surveys, as (i) they vary much less steeply with cell size than the $\bar{\omega}_J$, and (ii) the projection factors from the three-dimensional $S_J(r)$ to the two-dimensional $s_J(\theta)$ are almost constant with survey depth. 

As the $S_3$ and $S_4$ projection factors modelled by Szapudi et al. (1995) remain constant within 4 per cent for  3 mag shifts in survey limit,
we neglect any change in $s_J(\theta)/S_J(r)$ between the $17<B<20$ and $18.5\leq R\leq 23.5$ surveys to be negligable and plot the $s_{J}(\theta)$ from the shallower surveys without any vertical shift.
 However, 
we must take into account the difference in the mean proper distance $r$ corresponding to a given cell size $\theta$ in the deep and shallow data. Our PLE model gives a mean angular diameter distance of
$335 h^{-1}$ Mpc for $17\leq B\leq 20$ galaxies (with $z_{mean}=0.14$) and $860 h^{-1}$ Mpc for $18.5\leq R\leq 23.5$ galaxies (with $z_{mean}=0.69$), so we plot the APM and EDSGCS $s_{J}(\theta)$ shifted horizontally by $\Delta({\rm log} \theta)=\rm log {335\over 860}=-0.41$.  For all data on Figures 15 and 16, 
 log $\theta=-1$ will then correspond to, on average, $r\simeq 1.50 h^{-1}$ Mpc.
 
\begin{figure}
\psfig{file=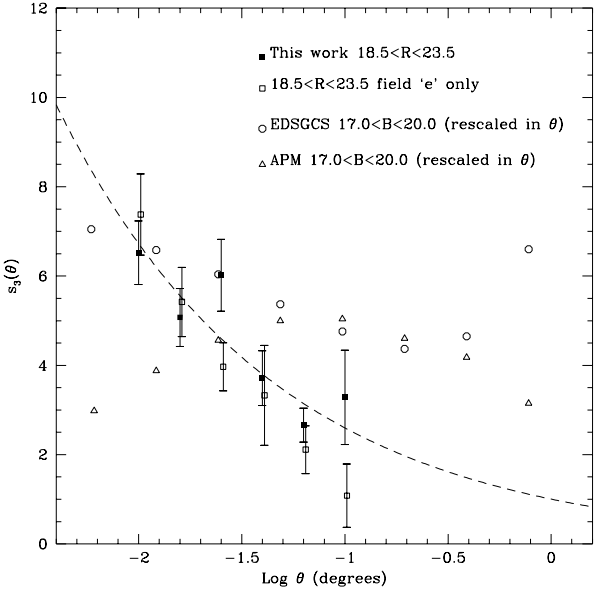,width=85mm}
\caption{The hierarchical moment $s_3(\theta)$ of our $18.5\leq R\leq 23.5$ 
galaxies, for both fields combined and for field `e' only (offset slightly for clarity), compared with the $s_3(\theta)$ of $17\leq B\leq 20$ galaxies in the APM (Gazta\~{n}aga 1994) and EDSCGS (Szapudi et al. 1996) surveys (shifted by
 $\Delta({\rm log} \theta)=-0.41$ to take into account the difference in survey depth). The dashed line shows the power-law ($\theta^{-0.413}$) best-fitting our results for the combined fields.} 
\end{figure}

\begin{figure}
\psfig{file=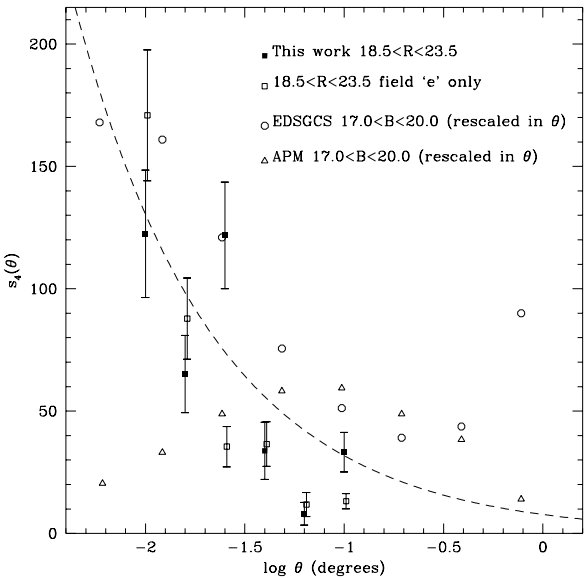,width=85mm}
\caption{The hierarchical moment $s_4(\theta)$ of our $18.5\leq R\leq 23.5$ 
galaxies, for both fields combined and for field `e' only (offset slightly for clarity), compared with the $s_4(\theta)$ of $17\leq B\leq 20$ galaxies in the APM (Gazta\~{n}aga 1994) and EDSCGS (Szapudi et al. 1996) surveys (shifted by
 $\Delta({\rm log} \theta))=-0.41$ to take into account the difference in survey depth). The dashed line shows the power-law ($\theta^{-0.614}$) best-fitting our results for the combined fields.} 
\end{figure}

Szapudi and Gazta\~{n}aga (1998) discuss the differences between the APM and 
EDSGCS $s_{J}(\theta)$. These are not corrected for integral constraint, resulting in
 spurious upturns at $\theta>3 $ deg for the APM survey and 
$\theta>1.5$ deg for the smaller-area EDSGCS survey, which account for the difference 
in these two  $s_{J}(\theta)$ at the largest $\theta$ plotted here. However, at $\theta<0.125$
 deg, here plotted as $\rm {log} (\theta)< -1.3$ and corresponding to $r<0.75 h^{-1}$ Mpc, the APM
and EDSGCS $s_{J}(\theta)$ are genuinely inconsistent, diverging greatly.

 Our results, for both $s_3(\theta)$ and $s_4(\theta)$, are clearly much more consistent with the steeper slopes of the EDSGCS 
moments over the same range of physical scales ($0.15\leq r\leq 1.5 h^{-1}$ Mpc). There is some indication that our $s_{J}(\theta)$ are lower in amplitude than those from
the EDSGCS, but the reduction is relatively small -- at $\rm log(\theta)\simeq 1.6$, the power-laws best-fitting our results  fall below the EDSGCS moments by
only $24\pm13 $ per cent for $s_3(\theta)$ and  $39\pm 18$ per cent for $s_4(\theta)$.  
We discuss the implications of the $s_J(\theta)$ slope and amplitude in Section 9.3.

\section {Discussion}

\subsection {Clustering Evolution}

Using a large sample of $\sim 70000$ galaxies on two fields,
 we have measured the angular
correlation function to a magnitude limit of $R=23.5$. 
 After 
corrections for star-contamination, our results support the relatively high
normalization of $\omega(\theta)$  from the Infante and Pritchet (1995) and Woods and Fahlman (1997) surveys, which used very high Galactic latitude fields
with minimal star-contamination.

We can also compare our $\omega(\theta)$ amplitudes with those from surveys
in  other passbands at limits with a similar galaxy surface density. Our $\omega(\theta)$ amplitudes at $R\sim 22.5$--23.5 are similar to those
measured at $I\sim 22$--24 (Neuschaefer et al. 1995; Brainerd and Smail 1998)
and $K\sim 19.5$--21.5 (Carlberg et al. 1997; Roche et al. 1998b)
but higher than at limits of $B\sim 24.5$ (Roche et al. 1993) and $V\sim 24.5$
(Woods and Fahlman 1997). This would be expected from the higher $\omega(\theta)$ amplitude of redder galaxies when the sample is 
divided by $V-I$ (Neuschaefer et al. 1995) $B-R$ (Roche et al. 1996) or $U-K$ (Carlberg et al. 1997).
 Neuschaefer et al. (1997) also find the
$\omega(\theta)$ amplitude of bulge-profile galaxies  at $I\leq 23$ to be 2--4 times higher than that of disk galaxies, suggesting that the stronger clustering of earlier Hubble types (LMEP, Guzzo et al. 1997) is maintained to at least $z\sim 0.7$. This might explain why the increase in  $\omega(\theta)$ towards longer survey  wavelengths appears to occur primarily between the $V$ and $R$ passbands, as these would lie on either side of the 
$4000\rm \AA$ break for galaxies near the peak of $N(z)$ at these limits, and the
size of this break correlates strongly with Hubble type.

Our results are consistent with the strength of galaxy clustering measured locally (LMEP), combined with galaxy $L^*$ evolution consistent with current 
redshift surveys, and galaxy clustering approximately stable ($\epsilon=0$)
in proper co-ordinates. Strong clustering
evolution of $\epsilon=1.2$, the linear model evolution of $\epsilon=0.8$, and comoving clustering ($\epsilon=-1.2$) are all disfavoured by $\sim 3\sigma$.
 The clustering simulations of
Col\'{i}n et al. (1997), for the distribution of mass, predict $\epsilon=1.08\pm 0.09$ with $\Omega=1$
and $\epsilon=0.18\pm 0.12$ with $\Omega=0.2$. Hence,
on the face of it, the observations 
appear to strongly favour a low
density Universe, but the situation is more complicated if 
the galaxy luminosity distribution is strongly biased relative to the
fluctuations in the density field. 

An extreme example of a biased model is the merging model of Matarrese et al. (1997), in which visible galaxies are identified with very massive ($\sim 10^{13}$ $h^{-1}M_{\sun}$) dark matter halos, with a linear bias relative to 
the mass fluctuations of $b=1.46$ at $z=0$ and evolving as $b=0.41+1.05(1+z)^{1.8}$. As $\xi(r)\propto b^2$, this evolution of biasing would from $z=0$ to $z=1$ increase $\xi(r)$ by a factor 7.76, producing 
clustering evolution of approximately $\epsilon_{galaxies}\simeq \epsilon_{mass}-3$.
Not suprisingly, this model overpredicts the observed faint galaxy $\omega(\theta)$ and could be rejected in any cosmology (Moscardini et al. 1998). 

Matarrese et al. (1997) also present a `transient model' in which star-forming faint galaxies are identified
with less massive ($\sim 10^{11}$ $h^{-1}M_{\sun}$) halos and are 
assumed to evolve into a weakly
clustered ($b=0.67$ hence $r_0\simeq 2$ $\rm h^{-1}Mpc$), low surface brightness population at the present day (see also Efstathiou 1995). The bias again evolves
rapidly, as $b=0.41+0.26(1+z)^{1.85}$, which at $z\sim 1$ gives a similar $\xi(r)$ to a model with normally clustered ($b\simeq 1$) galaxies and  $\epsilon\sim 0$. For a $z<1.6$ sample (approximately the depth of our data) the transient model $\omega(\theta)$ is indistinguishable from an unbiased ($b=1$) model, so our $\omega(\theta$) results alone may  not be sufficient to exclude the transient model. Higher-order correlation functions (Section 9.3)
may provide further constraints on galaxy biasing.

  Limits on clustering evolution from this data will
only apply to $z\sim 1.5$, as few galaxies will be detected
beyond this redshift
(Figure 11). The clustering of Lyman break galaxies at $z\sim 3$ (Giavalisco et al. 1998) is even stronger than expected from our $\epsilon=0$ model, indicating a negative $\epsilon$ and an increase in linear bias at these redshifts. 
However, 
at least for the $I$-limited HDF sample, this increase in bias appears to affect $\omega(\theta)$  only at $z>2.4$ (Magliocchetti and Maddox 1998), in contrast with the Matarrese et al. (1997) merging model in which bias rapidly increases at all $z>0$. As only $\sim 14$ per cent of galaxies  at the HDF limit are at $z>2.4$,
the $\omega(\theta)$ for the full $I$-limited sample (Villumsen et al. 1997) remains very low.

Villumsen et al. (1997) originally interpreted the low $\omega(\theta)$ amplitudes on the Hubble Deep Field as favouring $r_0\simeq 4$ $h^{-1}$ Mpc locally with $\epsilon=0.8$ clustering
evolution, whereas our results are much more consistent with $\epsilon=0$. Furthermore, faintward of $R\sim 26$, it is unlikely that $N(z)$ will become much more extended, and the most important change in $N(z)$ is that less luminous galaxies are seen in the same volume of space. Hence it seems more likely that any further decrease in $\omega(\theta)$ amplitude at $R>26$ is the
result of a decrease in the intrinsic strength of clustering
($r_0$) at low luminosities rather
than clustering evolution. This interpretation would be
supported by the  results of Connolly, Szalay and Brunner (1998), who 
for photometric-redshift-divided HDF galaxies at $I\leq 27$
 found weak intrinsic clustering of $r_0=2.37$ $h^{-1}$ Mpc but approximately stable clustering ($\epsilon=-0.4^{+0.37}_{-0.65}$) over the $0.4<z<1.6$ range   (where dwarfs will dominate the sample).

We find no evidence of a flattening of the slope of $\omega(\theta)$ 
on going faintward (Table 4).  None of  our calculated $\omega(\theta)$ are significantly flatter than $\delta=0.8$, at any magnitude limit, which suggests that the $\xi(r)$ of the spiral galaxies forming the greater part of our catalog is close to $\gamma=1.8$, and certainly no flatter than the $\gamma=1.72$ of LMEP.
Giavalisco et al. (1998) estimate a 
$\delta=0.98^{+0.32}_{-0.28}$ for  the $\omega(\theta)$ of $z\sim 3$ galaxies, suggesting
the relatively steep slope we observe is maintained to much higher redshifts.

A constant slope for $\omega(\theta)$ may not be surprising in view of the
 Col\'{i}n et al. (1997) simulations which predict no significant flattening of the mass distribution $\xi(r)$ , even to $z\sim 5$, for
any $\Omega$. They do
predict some flattening ($\Delta(\gamma)\simeq -0.2$ to $z\sim 1$)
of the galaxy $\xi(r)$ for a strongly biased model, but such models
 already seem disfavoured by the $\omega(\theta)$ scaling.  
Some flattening of $\omega(\theta)$  between $R=21$ and $R=23.5$
might be expected  from the decrease in the fraction of ellipticals (Figure 12b), but we estimate that this would only amount to $\Delta(\delta)\simeq -0.02$ to $-0.04$ (from the LMEP and Guzzo et al. $\gamma$ respectively),
and might be cancelled out by the increase in the fraction of lower luminosity latte-type galaxies, as
these may  also have a steeper $\xi(r)$ than $L^*$ spirals (LMEP estimate
$\gamma=2.01\pm0.1$).  

Further investigation of the evolution 
of $\omega(\theta)$
will require large-format CCD surveys of a similar area to a greater depth, with imaging in 
three or more passbands (preferably including a near-infra-red passband) to
enable a division of all the galaxies by both Hubble type and photometrically
estimated redshift.

\subsection {Merger Rate Evolution}

On the larger of our two fields, we find a $\sim 4\sigma$
significance excess in $\omega(\theta)$ at
$2\leq \theta\leq 5$ arcsec compared to the $\theta^{-0.8}$
power-law fitted at $\theta>5$ arcsec. This was interpreted in terms of the fraction of the total sample of galaxies within $\theta<5$ arcsec pairs in excess of the expectation from the clustering seen at larger
separations, which appeared to increase on going faintwards.
       The evolution in the  pair fraction will follow approximately the
evolution of the merger rate, parameterized here as $R_m=R_{m0}(1+z)^m$.

      By comparing the fraction of galaxies in close pairs at $V\leq 22.5$ with
that at very low redshifts, Carlberg et al. (1994) claimed rapid evolution of
$m=3.4\pm 1.0$, but this was based on a very small sample.
With a dataset $\sim 40$ times larger, Infante et al. (1996)
estimated a merger rate evolution of $m=2.2\pm 0.5$ at $R\leq 21.5$, and we similarly estimate   
  $m=2.01_{-0.69}^{+0.52}$
for our field `e' data to $R=23.5$. 

The Infante et al. (1995) results and ours are 
consistent with the $m\simeq 2$ expected for a low density ($\Omega\simeq 0.2$) Universe, but inconsistent with the $m\simeq 4$ predicted for $\Omega=1$ (Carlberg et al. 1994). Neuschaefer et al. (1997) derived an
even lower rate of $m=1.2\pm 0.4$ from the close pair fraction on deeper ($I\leq 25$) HST data, which may hint at a reduction in $m$ with increasing survey
depth, and would be consistent with the Carlberg et al. (1994) model for a low $\Omega$ Universe in which  merging leads to a moderate reduction in mean galaxy mass at high 
redshifts.

\subsection {Hierarchical Moments and Non-linear Bias}

Using a simple counts-in-cells method, we investigated the area-averaged
three and four point correlation functions of  $18.5\leq R\leq 23.5$ galaxies,
in the form of the hierarchical moments $s_3(\theta)$ and
$s_4(\theta)$. Both moments were detected with high statistical significance,
and, over a range of cell sizes $0.01\leq \theta \leq 0.1$ deg, showed 
relatively steep slopes best-fitted with power-laws $\theta^{-\delta}$ 
with $\delta=0.413\pm 0.083$ for $s_3(\theta)$ and  $\delta=0.614\pm 0.134$ for $s_4(\theta)$.
These slopes appear consistent with those of the EDSGCS $s_{J}(\theta)$ at equivalent scales 
($0.15\leq r\leq 1.5 h^{-1}$ Mpc), but strongly inconsistent with the near zero or positive $\delta$ of the APM $s_{J}(\theta)$.
 
 Szapudi and Gazta\~{n}aga (1998) attribute the difference between the APM and EDSGCS
 $s_{J}(\theta)$ at small scales to the more efficient deblending of
galaxy images in rich cluster fields in the EDSGCS analysis, but also claim that the EDSGCS contained some spurious detections, and so 
conclude that `the true galaxy distribution is likely to  
lie somewhere between the two surveys'. Our results, based on CCD 
data giving much more reliable deblending than digitized photographic plates, suggest the EDSGCS results are the more accurate.

Furthermore, a recent direct measurement of the three-point correlation
function of $0.03<z<0.15$ galaxies (Jing and B\"{o}rner 1998) in the Las Campanas Redshift Survey (also selected using CCD data)
 found the ratio of the projected three-point 
correlation function to $\xi(r)$ to scale with separation as
$r^{-0.3}$ at $0.2<r<3 h^{-1}$. These results were consistent with the slope and normalization of the EDSGCS $s_3(\theta)$, and with our best-fit slope for $s_3(\theta)$, but are inconsistent with the flat APM $s_3(\theta)$.

The $S_{J}(r)$
from $N$-body simulations 
(Baugh, Gazta\~{n}aga and Efstathiou 1995) do show some rise towards smaller 
cell sizes, which is steeper for successively higher order moments, as observed here and 
by Szapudi et al. (1996). The simulations give no suggestion that the
 $S_{J}(r)$ would flatten between $z_{mean}=0.69$ and $z_{mean}=0.14$ epochs, but predict a slight steepening with time, and so may strengthen the case that our results favour the EDSGCS $s_{J}(\theta)$ over the APM estimate. The predicted $S_{J}(r)$ are steeper for lower density cosmologies  (Gazta\~{n}aga and Baugh 1995); our results are more consistent with the $\Omega=0.2$ than the $\Omega=1$ models,
as is the three-point correlation function of Jing and B\"{o}rner (1998).

At this point many of our results ($\epsilon\simeq 0$ galaxy clustering, the moderate evolution of the merger rate, the steep $s_3(\theta)$ and $s_4(\theta)$) seem to favour a low $\Omega$ cosmology, but this interpretation assumes that galaxies at least
approximately trace the underlying mass distribution. 
Strongly biased models (e.g. Matarrese
et al. 1996; Moscardini et al. 1998) might be adjusted to fit the $\omega(\theta)$ scaling for a wide range of cosmological models, but, as we 
discuss below, higher-order correlation 
functions may have the potential
to break the apparent degeneracy between the biasing and cosmological model.

For a distribution of mass $\rho_{mass}(\bmath{r})$, with
fluctuations about the mean $\delta_{mass}(\bmath{r})=\rho_{mass}(\bmath{r})
-\langle \rho_{mass} \rangle$, and a distribution of the galaxy luminosity
in a particular passband  $\rho_{gal}(\bmath{r})$, with
fluctuations $\delta_{gal}(\bmath{r})=\rho_{gal}(\bmath{r})
-\langle \rho_{gal} \rangle$,
the biasing of galaxies relative to mass can be expressed as the series
(e.g Fry and Gazta\~{n}aga 1993)
\begin{equation}
\delta_{gal}(\bmath{r})=\sum_{k=0}^{\infty} {b_{k}\over k!} \delta_{mass}(\bmath{r})
\end{equation}

The galaxy $\xi(r)$ can be interpreted as the $\xi(r)$ of the underlying
mass distribution multiplied by $b_{1}^{2}$, where $b_{1}$ is the linear
bias. The hierarchical moments of the galaxy distribution are related to those
of the mass distribution through expressions involving the higher-order (non-linear) 
bias coefficients,
\begin{equation}
s_{3(gal)}=(s_{3(mass)}+3{b_2\over b_1})/b_1
\end{equation}
\begin{equation}
s_{4(gal)}=(s_{4(mass)}+12{b_2\over b_1}s_{3(mass)}+4{b_3\over b_1} 
+12 {b_2^2\over b_1^2})/b_1^2
\end{equation}
Gazta\~{n}aga and Frieman (1994) found the APM survey hierarchical moments to be consistent with perturbation theory predictions for  a simple model in which the blue-band luminosity from galaxies 
linearly traces the mass distribution -- i.e.
$b_{1}\simeq 1$  and $b_{k}\simeq 0$ for all $k>1$ -- with the best
 fit for 
a low density ($\Omega=0.2$ $\Lambda=0.8$) CDM model. Improved modelling with $N$-body simulations (Baugh, Gazta\~{n}aga and Efstathiou 1995; Gazta\~{n}aga and Baugh 1995) gave steeper $S_J(r)$ at $r\leq 7 h^{-1}$ Mpc for a given $\Omega$, but appears to give the same 
interpretation when compared with the EDSGCS results
(Szapudi and  Gazta\~{n}aga 1998).
 
 However, even if an optically-selected sample as a whole is approximately unbiased, the
different types of galaxy 
may vary greatly in biasing properties. For example, $60\rm \mu m$-selected IRAS galaxies, in addition to having a lower $\xi(r)$ amplitude than optically selected galaxies (implying $b_{1(IRAS)}\simeq 0.7$), are a factor of 2 lower in
$s_3(\theta)$ and a factor 3 lower in $s_4(\theta)$ (Fry and Gazta\~{n}aga 1993). This indicates that IRAS galaxies have a non-linear bias with
$b_{2}<0$, so that they tend to avoid the richer clusters in the galaxy distribution as a whole.

In simulations (Baugh, Gazta\~{n}aga and Efstathiou 1995; Gazta\~{n}aga and Baugh 1995), the $S_J(r)$
of the underlying mass distribution  
slowly increase and steepen with time at the $r\geq 7 h^{-1}$ Mpc distances 
relevant to our survey. At a fixed proper separation $r\sim 1 h^{-1}$
Mpc, the models appear to show a decrease of $\sim 20$--25 per cent for
 $S_3(r)$ and $\sim 30$--40 per cent for $S_4(r)$ on going from the mean epoch of the EDSGCS/APM galaxies to that of the $18.5\leq R\leq 23.5$ data. The decrease we observe is consistent with this model, implying that the  
 mean linear and non-linear biasing
properties of the luminosity from galaxies (in the rest-frame blue-band) galaxy remain approximately constant to $z\sim 1$.

We interpret this as evidence that (i) most rich galaxy clusters formed  prior to our mean redshift $z\sim 0.7$, as otherwise the galaxy distribution at these redshifts would be more like local IRAS galaxies, resulting in a further reduction of as much as a factor of $\sim 2$--3 in the $s_J(\theta)$ and (ii) galaxies
 with different bias properties (cluster and field galaxies, ellipticals and 
spirals,
 dwarf and giant galaxies) undergo a similar
evolutionary brightening in the rest-frame blue-band out to $z\sim 1$, as in PLE models. 

Our results 
 would argue against `transient' or `disappearing dwarf' models (e.g. 
Babul and Rees 1992; Efstathiou 1995), 
in which a large proportion of the faint blue galaxies are weakly clustered low
mass systems undergoing brief starbursts.
 Although the `transient model' of
 Matarrese et al. (1996) would be consistent with the $\omega(\theta)$ at 
$R=23.5$, the transient dwarfs are associated with low mass haloes collapsing at relatively late epochs, which would have a very negative
$b_2$ (Mo and White 1996). Domination of the deep sample by such objects would therefore change the nonlinear biasing properties relative to the APM/EDSGCS galaxies and give $s_3(\theta)$ and $s_4(\theta)$ lower than are observed.
PLE type evolution is also favoured by combined HST and spectroscopic data 
which  provide direct evidence that  field and
cluster galaxies (Schade et al. 1996a, 1996b) and early and late types (Roche et al. 1998;
Glazebrook et al. 1998) all undergo similar  blue-band evolution of
$\sim 1$ mag to $z\sim1$, so that, at least in visible-light wavelengths, the evolution required to fit the number counts is largely shared amongst normal galaxies rather than being
concentrated into a new type of object.

Furthermore, if galaxies, on average, linearly trace the mass distribution at both low redshifts and the depths of our survey, the faint galaxy $\omega(\theta)$ scaling indicates that clustering is stable ($\epsilon\simeq 0$) for the mass distribution and not just for the galaxy luminosity as observed in a particular passband.
Hence the hierarchical moments strengthen our earlier conclusions that galaxy clustering is genuinely stable to $z\sim 1$ and that we live in a low $\Omega$ Universe.

As larger areas are surveyed at CCD depths, the hierarchical moments
are likely to become an important complement to the study of the faint galaxy $\omega(\theta)$.  
 We hope to further these investigations, 
As for $\omega(\theta)$, by surveying
a similar area to a greater depth and by observing in three or more passbands, including one in the near-infra-red, so that colour-divided samples will
better constrain the evolution of $s_{J}(\theta)$ and show the relation between non-linear bias and galaxy colour.

\subsection*{Acknowledgements}

The Isaac Newton Telescope is operated on the island of La Palma by the
Isaac Newton Group in the Spanish Observatorio del Roque de los Muchachos
of the Instituto de Astrofisica de Canarias.  The data
reduction and analysis were carried out at the University of Wales 
Cardiff, using
facilities provided by the UK Starlink project,
funded by the PPARC.
Nathan Roche acknowledges the support of a PPARC research associateship.


\begin{thebibliography}{}
\bibitem{}
     Babul A., Rees M. J., 1992, MNRAS, 255, 346.
\bibitem{}
     Bahcall J. N., Soneira R. M., 1980, ApJS 44, 73.
\bibitem{}
     Baugh C. M., Gazta\~{n}aga, E., Efstathiou, G., 1995, MNRAS 274, 1049.
\bibitem{}
     Bernstein G. M., 1994, ApJ 424, 569.
\bibitem{}
     Bershady M., Majewski S., Koo D. C., Kron R. G., Munn J. A., 1997,
     ApJ, L41. 
\bibitem{}
     Brainerd T. G., Smail  I., Mould, J. R., 1995, MNRAS, 275, 781.   
\bibitem{}
     Bromley B. C., Press W. H., Lin H., Kirshner, R. P., 1998, ApJ, 505, 25.
\bibitem{}
     Burstein D., Heiles C., 1978. ApJ, 225, 40.
\bibitem{}
     Carlberg R. G., Cowie L. L., Songaila  A., Hu E. M., 1997. ApJ, 484, 538.
\bibitem{}
     Carlberg R. G., Pritchet C. J., Infante  L., 1994. ApJ, 435, 540.
\bibitem{}
     Charlot S., Worthey G., Bressan A., 1996, ApJ, 457, 625.
\bibitem{}
     Col\'{i}n P., Carlberg R. G., Couchman H. M. P., 1997. ApJ, 490, 1.
\bibitem{}
     Connolly A.J., Szalay A.S., Brunner R.J., 1998, ApJ, 499, L125.
 \bibitem{}
     Couch W. J., Jurc\'{e}vic J. S.,  Boyle B. J., 1993, MNRAS, 260, 241.
\bibitem{}
     Cowie L. L., Songaila A., Hu, E. M., 1996, AJ, 112, 839.
\bibitem{}
     Efstathiou G., 1995, MNRAS, 272, L25.
\bibitem{}
     Efstathiou G., Bernstein G., Katz N., Tyson J.A., Guhathakurta  P.,
     1991, ApJ, 380, L47.
\bibitem{}
     Fry J. N., Gazta\~{n}aga, E. 1993, ApJ, 413, 447.
\bibitem{}
     Gazta\~{n}aga E., 1994, MNRAS, 268, 913.
\bibitem{}
     Gazta\~{n}aga E., Baugh C. M., 1995, MNRAS, 273, L1.
\bibitem{}     
     Gazta\~{n}aga E., Frieman J., 1994, ApJ, 437, L13.
\bibitem{}
     Giavalisco M., Steidel C. C., Adelberger K. L., Dickinson M., Pettini M.,
     Kellogg M., 1998, ApJ, 503, 543.  
\bibitem{}
     Groth E. J., Peebles P. J. E., 1977, ApJ, 217, 385.
\bibitem{}
     Guzzo L., Strauss M. A., Fisher K., Giovanelli R., Haynes M., 1997,
     ApJ, 489, 37. 
\bibitem{}
     Infante L., de Mello D. F., Menanteau, F., 1996, ApJ, 469, L85.
\bibitem{}
     Infante L., Pritchet C.J., 1995, ApJ, 439, 565.
\bibitem{}
     Jing Y. P., B\"{o}rner G., 1998, ApJ, 503, 37.
\bibitem{}
     Landolt A. U., 1992, AJ, 104, 340.
\bibitem{}
     Landy S. A., Szalay A.S., 1993, ApJ, 412, 64.
\bibitem{}
     Le F\`{e}vre O., Hudon D., Lilly S. J., Crampton D., Hammer F., 
     Tress\`{e} L., 1996, ApJ, 461, 534. 
\bibitem{}
     Loveday J., Maddox S. J., Efstathiou G., Peterson B. A., 1995, ApJ,
     442, 457.
\bibitem{}
     Madau P., Ferguson H., Dickinson M., Giavalisco M., Steidel C.,
     Fruchter A., 1996, MNRAS, 283, 1388.
\bibitem{}
     Maddox S. J., Efstathiou G., Sutherland W. J., Loveday J., 1990,
     MNRAS, 242, L43.
\bibitem{}
     Magliocchetti M., Maddox S.J., MNRAS, submitted (astro-ph/9811320). 
\bibitem{}
     Marzke R. O., Geller M. J., Huchra J. P., Corwin  H. G. Jr., 1994, AJ, 
     108, 437. 
\bibitem{}
     Matarrese S., Coles P., Luccin F., Moscardini L., 1996, MNRAS 286, 115.
\bibitem{}
     Mathis J., 1990, A\&A Ann. Rev, 38, 37.
\bibitem{}
     Metcalfe N., Shanks T., Fong R., Jones L., 1991, MNRAS, 249, 498.
\bibitem{}
     Metcalfe N., Shanks T., Fong R., Roche N., 1995, MNRAS, 273, 257.
\bibitem{}
     Mihos J. C., Hernquist L., 1996, ApJ, 464, 641.
\bibitem{}
     Mo, H. J., White, S.D.M., 1996, MNRAS 282, 347.
\bibitem{}
     Moscardini L., Coles P., Lucchin F., Matarrese S., 1998, MNRAS, 299, 95.
\bibitem{}
     Neuschaefer L. W., Ratnatunga K., Griffiths R.E, Casertano S., 1995,
     ApJ, 453, 559. 
\bibitem{}
     Neuschaefer L. W., Ratnatunga K., Griffiths R.E, Casertano S., 1997,
     ApJ, 480, 59. 
\bibitem{}
     Phillipps S., Fong R., Ellis R. S., Fall S. M., MacGillivray H. T., 1978,
     MNRAS, 182, 673.
\bibitem{}
     Ratcliffe A., Shanks T., Parker Q. A., Fong R., 1998, MNRAS, 296, 173.
\bibitem{}
     Reid J. N., Yan L., Majewski S., Thompson I., Smail I., 1996, AJ, 112, 1472.
\bibitem{}
     Roche N., Eales S., Hippelein H., 1998b, 295, 946.
\bibitem{}
     Roche N., Ratnatunga K., Griffiths R. E., Im M., Naim A., 1998a,
     MNRAS 293, 157. 
\bibitem{}
     Roche N., Shanks T., Metcalfe N., Fong R., 1993, MNRAS, 263, 360.
\bibitem{}
     Roche N., Shanks T., Metcalfe N., Fong R., 1996, MNRAS, 280, 397.
\bibitem{}
     Schade D., Carlberg R. G., Yee H. K. C., Lopez-Cruz O., Ellingson E.,
1996a, ApJ, 464, 63.
\bibitem{}
     Schade D., Carlberg R. G., Yee H. K. C., Lopez-Cruz O., Ellingson E.,
1996b, ApJ, 465, 103.
\bibitem{}
     Shepherd C. W., Carlberg R. G., Yee H. K. C., 1997, ApJ, 479, 82.
\bibitem{}
     Steidel C. C., Giavalisco M., Dickinson M., Adelberger K. L., 1996,
     AJ, 112, 352.
\bibitem{}
     Szapudi I., Dalton G., Efstathiou G., Szalay A., 1995, ApJ 444, 520.
\bibitem{}
     Szapudi I., Gazta\~{n}aga E., 1998, MNRAS, 300 , 493.  
\bibitem{}
     Szapudi I., Meiksin A., Nichol R., 1996, ApJ, 473, 15. 
\bibitem{}
     Villumsen J. V., Freudling W., da Costa, L., 1997. ApJ, 481, 578.
\bibitem{}
     Woods D., Fahlman, G., 1997, ApJ, 490, 11.

\end{thebibliography}
\end{document}